\documentstyle[psfig,aps]{revtex}
\begin{document}
\title{
ELECTRON-BOSON EFFECTS IN THE INFRARED 
 PROPERTIES OF METALS}
%\subtitle{What every metal worker has to know about 
  %FIR properties of metals }
%\subtitle{What can we learn from the optical conductivity \vspace{-4.4cm} 
%$\mbox{\ \ \ \ \ \ \ \ \ \ \ \ \ \ \ \ \ \ \ \ \ \ \ \ 
%\ \ \ \ \ \ \ \ \ \ \ \ \ \ \ \ \ \ \ \ \ \ \ \ \ \ \ \ \ \ \ \ \ \ \ \ \ \ \ \ \ \ \ \ \ \ \ \ 
%\ \ \ \ \ \ \ \ \ \ \ \ \ \ \ \ \ \ \ \ \ \ \ \ \ \ \ \ }$What 
\author{S.\  V.\  SHULGA}
\address{Institute of Spectroscopy,  RAS, Troitsk, 142190, Russia, \\
IFW-Dresden and the University of Bayreuth, Germany}
%\runningtitle{ELECTRON-BOSON EFFECTS ...}
\date{September 1998}
\maketitle
\begin{abstract}
The interpretation of  optical conductivity in the normal and superconducting states 
 is considered in the frame of the standard
 Isotropic Single wide Band (ISB) model using the theory proposed by 
 Nam (Nam S.B., {\it Phys. Rev.}{\bf\ 156} 470-486 (1967)).
The {\it exact analytical} 
 inversion of the normal state Nam equations  is performed and applied
 for the recovery of the reliable information from the experimental data. 
The Allen formula
 is derived in the strong coupling approximation and used for the
 physically transparent interpretation of the FIR absorption. 
 The phenomenological ''generalised`` Drude formula is obtained from 
 the Nam  theory in a high temperature approximation. 
It is shown, that the reconstruction 
 of the shape of the spectral function $\alpha^2F(\omega)$
 from the normal state optical data
 at T$>$T$_c$ is not unique and the same
 data can be fitted by  many spectral functions. This problem is
considered in detail from  different  points of view.   
 At the same time, using the exact analytical solution, 
one can get from the normal state  data a useful piece of information, namely, 
 the value of the coupling constant, the upper energy bound of the electron-boson
interaction function, and the averaged boson frequency. Moreover,
they are even visually accessible, if the optical data are presented in term of 
the frequency dependent optical mass and scattering rate. 
The superconducting state Nam formalism and related simplified theory
are analysed from the user point of view. 
   A novel {\it adaptive} 
 method of 
 the {\it exact  numerical inversion} of the superconducting state 
 Nam equations is presented. Since this approach uses the
{\it first derivative}  from the experimental curve, the 
signal/noise ratio problem is discussed in detail.
 It is shown that,  the fine structure of the spectral function can be recovered 
 from  optical data in the case of {\it s-}wave pairing symmetry.
 In contrast, in the  {\it d-}wave case the resulting 
  image is approximately the convolution of the input electron-boson interaction function 
  and the Gauss distribution exp($-\omega^2/\Delta_{opt}^2$).  
  A simplified visual accessibility (VA) procedure is proposed for ''by eye`` analyses
 of the superconducting state optical reflectivity of the ISB metals
 with  {\it s} and {\it d} pairing symmetry. The bosons, responsible for the 
 superconductivity in YBa$_2$Cu$_3$O$_7$, exhibit a ''phonon-like``   spectral function
 with the upper frequency bound less than  500 cm$^{-1}$ and the averaged
 boson frequency
 near 300 cm$^{-1}$.
\end{abstract}

%\footnote{Invited Lecture for Nato-ASI ''Material Science, Fundamental Properties
%and Future Electronic Applications of High-Tc Superconductors``, 
%Albena, Bulgaria, September 1998, Kluwer Academic Publishers, Dortrecht,
%(2001), pp. 323-360\vspace{2.6cm}}

\section{Introductory remarks, reservations and basic equations}
\begin{itemize}
\item This is the preliminary version of invited lecture 
for Nato-ASI ''Material Science, Fundamental Properties
and Future Electronic Applications of High-Tc Superconductors``, 
Albena, Bulgaria, September 1998, Kluwer Academic Publishers, Dortrecht,
pp. 323-360 (2001) 
\item
The interpretation of the normal and superconducting state 
 Far Infrared Region (FIR) properties of metals is still rather ambiguous. In this
lecture I will restrict myself to the consideration of the 
 of  Isotropic Single wide Band (ISB) model based on the Nam formalism \cite{nam67}. 

\item  
From the optical conductivity $\sigma(\omega)$ one  derives 
the so called transport spectral 
function $\alpha^2_{tr}F(\omega)$.
It has 
the same spectral structure, i.e.\ the same number of peaks with similar relative positions 
 (see Fig.4 in \cite{allen71} and Fig.1 in \cite{farn76})
as a standard  $\alpha^2F(\omega)$ which  enters the  Eliashberg equations.
 But their  amplitudes can  be lower, and as a result
the transport coupling constant $\lambda_{tr}$ is less than  the standard  $\lambda $.
In this lecture I will not distinct the  $\alpha^2F(\omega)$ and  
$\alpha^2_{tr}F(\omega)$.  
\item If the electronic band 
is very narrow, the forward scattering could dominate. In this case the optical
properties manifests themselves by the negligible coupling, while the tunnelling spectroscopy 
 gives  reasonable finite coupling strength.
A similar problem arises when the quasiparticles from the flat or 
nested parts of the Fermi surface 
are mainly responsible for the superconductivity. 
Since they have  small Fermi velocities,
their influence  in  the transport response function 
which is proportional to $v_F^2$  is weak. Such a scenario was recently
 proposed for borocarbides \cite{shulga98}. 
\item In the frame of the Eliashberg theory it is impossible
to elucidate  the nature of the bosons which are responsible for the 
 superconductivity. In this context I will use the term boson,
 keeping  for short the standard electron-phonon notation  $\alpha^2F(\omega)$.

\item  I will restrict myself to the consideration of  the local (London) limit case,
 that is, to the normal skin effect scenario. The generalisation 
of results obtained to the  nonlocal (Pippard) limit is possible but 
will not be considered here due to the lack of space.

\item The power of the proposed methods  is illustrated analysing  available  experimental
data for a single-domain YBa$_2$Cu$_3$O$_{7-d}$ crystal \cite{renk92}.
Note, that to the best of my knowledge,
the  experimental HTSC superconducting state reflectivity was not been
 considered  from the inverse problem viewpoint at all, despite the single early attempt
\cite{sulewski87}.

\item Theoretical analysis of any physical quantity, as optical conductivity,
is performed  in frame of  some model which includes a theoretical formalism and some 
assumptions about the band structure, the anisotropy of the coupling function and so on.
The considered here ISB model is the standard default. 
Nevertheless, in frame of its {\it s-}wave version
the question: ''What is the value of the gap? `` has an unique answer, but for {\it d-}wave
case  the correct questions are either :''What are the values of the gaps? `` or
''What is the value of the maximum gap? ``. In other words, the absolute values and  
the physical meaning of the quantities are model dependent.
\item   If the model is reach enough,
one can calculate many physical properties using the same 
{\it set of  input (material) parameters}. A solution of the
 inverse problem  means 
  the determination of these material
 parameters from the experimental data.
A short list of the ISB model  input parameters is given in 
 section \ref{isbsub}. 
\item As a rule, the inverse problem is  ill-posed.
The experimental data contain less information, than we would like to know.
The reasons for are  the temperature induced broadening and the  finite 
signal to noise ratio S/N.
\item
In this lecture we will mainly be interested
 in the amount of the remaining information about the spectral
function  $\alpha^2F(\omega)$, which can be obtained 
from the optical data. 
In spectroscopy the unit of information is a peak. A single peak 
needs four parameters for its description: the upper and the low frequency bounds
$\Omega_{max}$ and  $\Omega_{min}$, the  position of the maximum and its amplitude.
An narrow symmetric peak within the broad band needs three parameters:
 the  amplitude, the halfwidth,
 and the position. The sum of all peaks yields the upper and lower frequency
bounds  $\Omega_{max}$  and $\Omega_{min}$ of the spectral band.
Since the  $\Omega_{min}$ of the $\alpha^2F(\omega)$ is usually equal to 
zero, the remaining three band parameters of  $\alpha^2F(\omega)$
can be determined even from the 
normal state data \cite{shulga91}. 
\end{itemize}

\subsection{Isotropic single band model}
\label{isbsub}
   The standard ISB model \cite{carbotte90} is the most developed part of the 
modern theory of superconductivity. It 
describes {\it quantitatively} the renormalization of
the physical properties of metals  due to the 
electron-boson   interaction.
The input  quantities  of the ISB model are
the density of states at the Fermi energy $N(0)$, the Fermi velocity
$v_F$, the  impurity scattering rate $\gamma_{imp}$, the Coulomb
pseudopotential
$\mu^{*}$, and the spectral function  $\alpha^2F(\omega)$
 of the  electron-boson interaction.   The scale of  transport and
optical properties is fixed by the plasma frequency
  $\hbar \omega_{pl}=\sqrt{4\pi e^2 v_F^2 N(0) /3}$.
In this section I present only the final expressions of the Nam theory, more 
detailed information can be found in \cite{nam67,allen71,lee89}. 

 In order to calculate any physical property in the superconducting state 
one needs a solutions of Eliashberg equations (EE) \cite{allen82}.
In my opinion, the following
EE representation is most suitable for numerical solution by iterations
\begin{eqnarray}
Im\tilde{\Delta}(\omega)&  = & \frac{i\gamma_{imp}}{2} 
\frac{\tilde{\Delta}(\omega)}
{\sqrt{\tilde{\omega}^2(\omega)-\tilde{\Delta}^2(\omega)}}+
\frac{\pi}{2}\int dy
\alpha^2F(\omega -y)
\nonumber \\
 &   &
\left [ \coth{\left(\frac{\omega -y}{2T}\right )}
-\tanh{\left ( \frac{y}{2T} \right) }\right]
\mbox{Re}\frac{\tilde{\Delta}(y)}{\sqrt{\tilde{\omega}^2(y)
-\tilde{\Delta}^2(y)}}
 \label{eq1} \\
Im\tilde{\omega}(\omega) &  = &
\frac{i\gamma_{imp}}{2}\frac{\tilde{\omega}(\omega)}
{\sqrt{ \tilde{\omega}^2(\omega)-\tilde{\Delta}(\omega) }}+
\frac{\pi}{2}\int dy
\alpha^2F(\omega -y),
\nonumber \\
 &   &
\left [ \coth{\left(\frac{\omega-y}{2T}\right )}
-\tanh{\left ( \frac{y}{2T} \right) }\right]
\mbox{Re}\frac{\tilde{\omega}(y)}{\sqrt{\tilde{\omega}^2(y)-\tilde{\Delta}^2(y)}},
\label{ImEl}
\end{eqnarray}
where 
 $\tilde{\Delta}(\omega)$ and $\tilde{\omega}(\omega)$ are the renormalized 
gap function and the renormalized 
frequency respectively, and 
$\gamma_{imp}$ denotes  the impurity scattering rate within  the Born approximation.
The real and the imaginary 
parts of the Eliashberg functions $\tilde{\Delta}(\omega)$ and $\tilde{\omega}(\omega)$
are  connected by  the Kramers-Kronig relations.  Hence, they  have the same
 Fourier images.  This reasoning yields  the fast  solution procedure.
 The convolution type integrals (\ref{eq1}-\ref{ImEl})
should be calculated by the Fast Fourier Transform (FFT) algorithm.
The inverse {\it complex} Fourier transformations of the results obtained
 give  {\it complex} values of 
$\tilde{\Delta}(\omega)$ and $\tilde{\omega}(\omega)$.   
If one would like to use this efficient FFT method also for
{\it d-}wave pairing symmetry, the process of arithmetic-geometric mean
\cite{abram70} for the evaluation
of the complex elliptic integrals ( see below Eqs. (\ref{nscm1},\ref{nscm2}) is strongly
recommended. The density of states Re$N(\omega)$ and the
density of pairs Re$D(\omega)$ 
\begin{eqnarray}
N(\omega) & = & \frac{\tilde{\omega}(\omega)}
{\sqrt{\tilde{\omega}^2(\omega)-\tilde{\Delta}^2(\omega)} },
\label{Notom} \\
D(\omega) & = & \frac{\tilde{\Delta}(\omega)}
{\sqrt{\tilde{\omega}^2(\omega)-\tilde{\Delta}^2(\omega)} },
\label{Dotom} 
\end{eqnarray}
could be approximated 
by  step functions. The convolution of any spectrum with the 
step function sign$(y^2-\Delta^2_0)$ results in the shift 
by $\Delta_0$ at $y>0$ and by $-\Delta_0$ at $y<0$ and its integration 
over y.  It is important, that due to the presence of singularities in $N(\omega)$ and $D(\omega)$
at $\omega=\Delta_0$ the more  complicated functions such as 
$\int d\omega^\prime N(\omega^\prime)N(\omega-\omega^\prime)$   and 
$\int d\omega^\prime D(\omega^\prime)D(\omega-\omega^\prime)$   still
contain a jump (now at $\omega=2\Delta_0$). 
  Since the convolution is a quite general
property of the Green functions approach,  one can expect that the first derivatives 
from a spectrum (of different nature) over frequency will reproduce 
the input spectral function $\alpha^2F(\omega)$ (may be with the appropriate 
phase distortion).  When applied to the optical conductivity, 
this approach gives visual accessibility (VA) procedure
discussed in the sec. 
\ref{super}.

  In the normal state the set (\ref{ImEl}) is reduced to the following 
simple formula
\begin{equation}
Im\tilde{\omega}(\omega)  = \frac{\gamma_{imp}}{2}+ \frac{\pi}{2}\int dy
\alpha^2F(y)
\left [ \coth{\left(\frac{y}{2T}\right )}
-\tanh{\left ( \frac{\omega-y}{2T} \right) }\right],
\label{nel}
\end{equation} 
which however looks unwieldy in comparison with the same formula
presented in Matsubara formalism
\begin{equation}
\tilde{\omega}(i\omega_n)\equiv \tilde{\omega}_n= \tilde{\omega}_{n-1}+2\pi T(1+\lambda_n)
\label{ommats}
\end{equation}
where
\begin{equation}
\lambda _{k}\equiv\alpha^{2}F(i\nu_k)=2\int dz\alpha^{2}F(z)z/
({z^{2} + \nu_k^2})
\label{aaa2}
\end{equation}
are kernels of the spectral function,
 $\omega_k=\pi T(2k+1)$ and
 $\nu_k=2\pi Tk$ are fermion ($\omega_k$) and boson ($\nu_k$) Matsubara
energies, respectively,
and
$\tilde{\omega}_0 = \gamma_{imp}/2+\pi T(1+\lambda_0)$.

The  normal state ISB optical conductivity
takes the following forms 
\begin{equation}
\sigma(\omega,T)=\!\frac{\omega^2_{pl}}{8\pi i  \omega}\int d y
\frac{
\mbox{\ tanh} \left (  \frac{y+\omega}{2T}\right )
-\mbox{\ tanh} \left (  \frac{y}{2T}\right )}
{\tilde{\omega}(\omega+y)-\tilde{\omega}^{*}(y),
}
\label{signr}
\end{equation}
and 
\begin{equation}
\sigma(i\omega_n,T)\equiv\sigma_n=\!\frac{\omega^2_{pl}}{4\pi n}\sum_{k=0}^{n-1}
\frac{1}{\tilde{\omega}_k+\tilde{\omega}_{n-k-1}},
\label{signm}
\end{equation}
in the real and imaginary axes  techniques, respectively.

The  tedious expression 
for  the superconducting state ISB optical conductivity has the form  
\cite{nam67,lee89,dolgov90}
\begin{eqnarray}
\sigma(\omega)& = & \frac{\omega_{pl}^2}{16\pi\omega}
\int dx \frac{g_{rr}\tanh{(x/2T)}}{\sqrt{\tilde{\Delta}^2(x)
-\tilde{\omega}^2(x)}+\sqrt{\tilde{\Delta}^2(x+\omega)
-\tilde{\omega}^2(x+\omega)}} \nonumber \\
& & -\frac{g_{rr}^{*}\tanh{[(x+\omega)/2T]}}{\left(\sqrt{\tilde{\Delta}^2(x)
-\tilde{\omega}^2(x)}+\sqrt{\tilde{\Delta}^2(x+\omega)
-\tilde{\omega}^2(x+\omega)}\right)^*} \nonumber \\
& & +\frac{g_{ar}\left\{\tanh{[(x+\omega)/2T]}-\tanh{(x/2T)}\right\}}
{\sqrt{\tilde{\Delta}^2(x)
-\tilde{\omega}^2(x)}+\left(\sqrt{\tilde{\Delta}^2(x+\omega)
-\tilde{\omega}^2(x+\omega)}\right)^*},
\label{sigsup} 
\end{eqnarray}
where 
\begin{eqnarray}
g_{rr} & = & 1-N(x)N(x+\omega)-D(x)D(x+\omega),  \nonumber \\
g_{ra} & = & 1+N^*(x)N(x+\omega)+D^*(x)D(x+\omega)
\label{cohfac}
\end{eqnarray}
are {\it coherent factors}.

\section{Normal state optical conductivity}

\subsection{The exact solution of an inverse problem}
\label{exact}
      
It is amusing, that the  Eq.\ (\ref{signm}) is  solvable 
 for $\tilde{\omega}_k$ and
Eq.\ (\ref{ommats}) for  $\lambda_k$. Hence, at least theoretically, an 
exact analytical solution exists. It takes three or four steps as  shown below.

 I) One has to perform the analytical continuation
of the conductivity $\sigma(\omega)$ from the real energy  axis, where we are living, to the
poles of the Bose distribution function (Matsubara boson energies)
$i\nu_k=2 \pi i T k$. This procedure is similar  to 
the Kramers-Kronig analysis, which is nothing other than the 
 analytical continuation from the real frequency  axis to itself.  
 For example, if one fits the data by 
a sum of Drude and  Lorentz terms or by formula (\ref{padx}),
 one has simply to  substitute the complex 
values  $i\nu_k$ for the frequency. Note,  that
 the  non-metallic 
(IR active direct phonon contribution and interband transition) part of the dielectric
permeability $\epsilon_{ph}(\omega)$  have to be subtracted
from the total one $\epsilon(\omega)$  before starting of the  analysis. In the genuine
far infrared region one could use the real constant 
$\epsilon_{\infty}=\epsilon_{ph}(0)$,
 instead of   
$\epsilon_{ph}(\omega)$, but if the spectral range is wide, the subtraction of 
 $\epsilon_{ph}(\omega)$ is not trivial.

II) The renormalized  frequencies   should be calculated from (\ref{signm}) 
as follows
\begin{eqnarray}
\tilde{\omega}_0&  = & \frac{\omega_{pl}^2}{8 \pi \sigma_1}, 
\label{esom1} \\
\tilde{\omega}_1&  = & \frac{\omega_{pl}^2}{4 \pi \sigma_2}-\tilde{\omega_0},
\label{esom2} \\
\tilde{\omega}_{n-1}&  = & \frac{\omega_{pl}^2}{2 \pi A_n}-\tilde{\omega_0},   
\mbox{where\ }A_n=\sigma_n-
\!\frac{\omega^2_{pl}}{4\pi n}\sum_{k=1}^{n-2}
\frac{1}{\tilde{\omega}_k+\tilde{\omega}_{n-k-1}}.
\label{esom3}
\end{eqnarray}

III) Inverting  Eq.\ \ (\ref{ommats}), one could evaluate
 the values of the spectral function $\lambda_n=\alpha^2F(\nu_n)$ we looked for 
\begin{eqnarray}
\lambda_0 &  = & \frac{ \tilde{\omega}_0-\gamma_{imp}/2 }{ \pi T} -1
\label{eslam1} \\
\lambda_n &  = & \frac{ \tilde{\omega}_n-\tilde{\omega}_{n-1} }{2 \pi T} -1.
\label{eslam2}
\end{eqnarray}

IV)  Using, for example, Pade polynoms \cite{Pade} one has to continue 
the electron-boson interaction function back to the  real axis.

\begin{figure}[t]
\centerline{\hbox{   
\hspace{-2.3cm} 
\psfig{figure=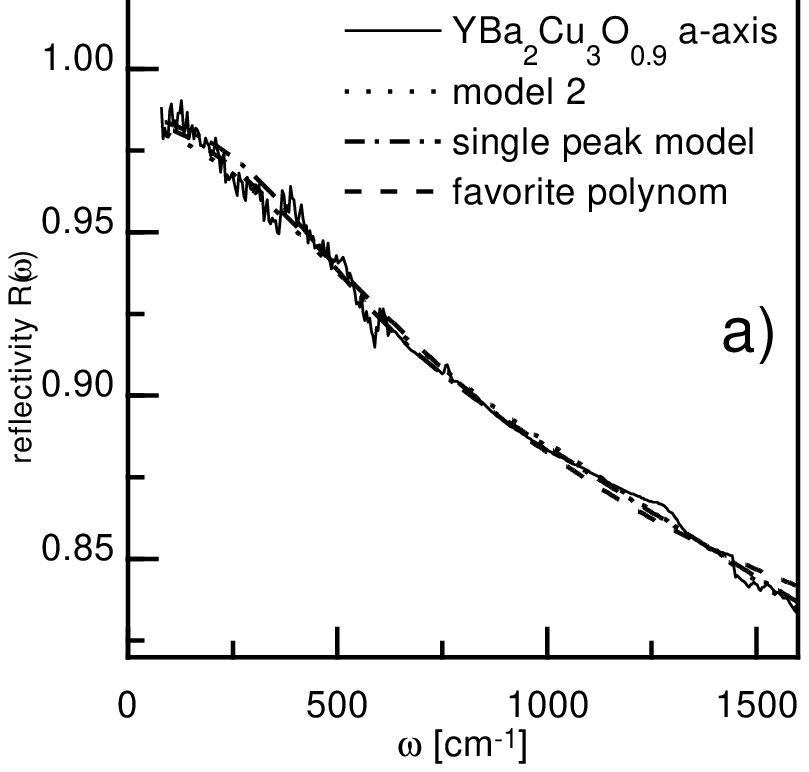,width=6.0cm,height=6.0cm}
\hspace{0.8cm} 
\psfig{figure=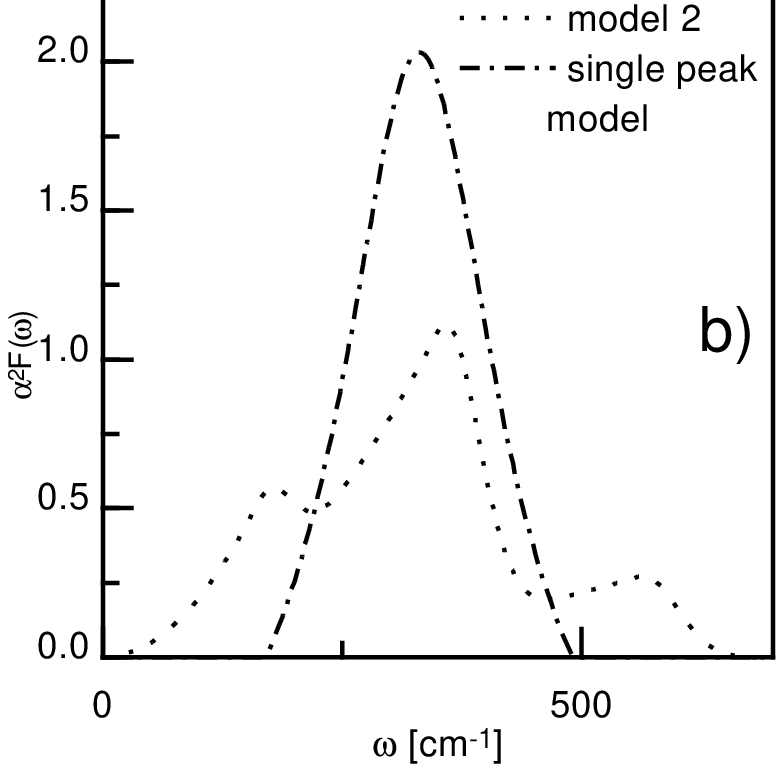,width=6.0cm,height=6.0cm}
}}
\vspace{1cm}
\caption{Panel (a) shows the  frequency dependence
 of the normal state  reflectivity 
(solid line, T=100 K, Schutzmann J. {et al.,} {\it Phys. Rev.}, {\bf B46}, 512 (1992))
 in comparison 
with the model ones.  The curves, shown by the dotted and the dotted-dashed lines,
was calculated  using ''model 2`` and ''single peak`` spectra (presented on the panel (b)).
The dashed line is the least square fit of YBa$_2$Cu$_3$O$_{0.9}$ a-axis
reflectivity by the favourite polynom Eq.\ \ 32 (with the fixed $\epsilon_{\infty}=12$).}
 \label{josfig}
 \end{figure}

In practice, if the signal to noise ratio is not extremely large,
the last step makes no sense due to the dramatic increase
of the uncertainty in the $\lambda_n$ with elevation of n.  The point is that 
the  small quantities  $\lambda_n$  Eq.\ \ (\ref{eslam2}) and $A_n$ Eq.\ (\ref{esom3})  are
the difference of  large quantities.  Note, that the  uncertainty in the $\lambda_n$ 
values has to be small, not in comparison with   $\lambda_n$ itself, but  with
the difference between  $\lambda_n$ and its high energy asymptotic value
\begin{equation}
\tilde{\lambda}_n=\frac{2\lambda_0<\Omega^2>}{<\Omega^2>+\nu_n^2}
\end{equation}
where
\begin{equation}
<\Omega^2>=\frac{1}{\lambda_0}\int_0^{\infty}d\Omega
\mbox{\ } \Omega\alpha^2F(\Omega)
\end{equation}
If the average boson energy has the order of $2\pi T$, 
the last  condition 
($\delta\lambda_n\ll|\tilde{\lambda}_n-\lambda_n|$) may
be violated even for n=3, since $\tilde{\lambda}_n$ and $\lambda_n$
have similar values.
  For example, at $T=$100 K for the 
spectral function, shown by the dotted line in Fig.\ \ref{josfig}, \ 
$\tilde{\lambda}_n/\lambda_n$=1.37, 1.12, 1.06, 1.03 for 
n=1, 2, 3, 4 respectively. Three or four $\lambda_n$ values only form a too small 
parameter set to recover correctly the shape of the spectral function.  
Actually, the possibility
to obtain at least  small amount of  reliable information,  is  the additional advantage
provided  by  the Matsubara solution in comparison with the real axis one where
{\it all} values are unreliable \cite{jos1}. 
Fig.\ \ref{josfig} shows the good agreement between 
the experimental  normal state 1-2-3 single crystal reflectivity
data \cite{renk92} and the calculated ones using  for
{\it  various } spectral functions.  
Unfortunately, the lack of a unique solution promoted   unreasonable
 speculations on the mechanisms in several novel superconducting systems in connection
with proposed hand-made electron-boson interaction functions.
In my opinion, twelve years after the discovery of the superconductivity 
in cuprates it is  timely to return to a self-consistent description of the 
superconductivity in this compounds.

  The first Matsubara value of the conductivity $\sigma(i\nu_1)(\equiv\sigma_1$) 
gives the value of the coupling 
constant $\lambda(\equiv\lambda_0)$, if  the 
plasma frequency $\omega_{pl}$ and the impurity scattering rate 
$\gamma_{imp}$ are known {\it a priori}
\begin{equation} 
 \frac{1}{ \sigma_1}=
 \frac{8 \pi^2 T(1+\lambda_0) }{\omega_{pl}^2}+\frac{4\pi \gamma_{imp}}{\omega_{pl}^2}.
\label{sig1}
\end{equation} 
 The reflectivity R$(\omega)$ can also  be continued to the imaginary frequency axis
and included into the solution of the inverse problem. Following this way, 
one can derive from
(\ref{sig1})  the following formula
\begin{equation}
\lambda_0=-\frac{\gamma_{imp}}{2\pi T}\ -1 \ +
\frac{\omega^{2}_{p}}{(2\pi T)^2(\coth^2(K_1/4)-\epsilon_{\infty})},
\label{lamR}
\end{equation}
where  $ \epsilon_{\infty}$ is the low-frequency value of the non-metallic 
(phonon and interband transition) part of dielectric
permeability,  and K$_m$ is the electromagnetic kernel
\begin{equation}
K(\nu_m)=\frac{2\nu_m}{\pi}\int_{0}^{{\infty}}dz
\frac{\log|1-R(z)|}
{z^2+\nu_{m}^{2}}.
\label{ImRm}
\end{equation}

 The first  $\lambda_1$ and the second $\lambda_2$ 
allow ( for example ) to estimate the average boson energy and {\it very approximately}
the width of  the spectral function.   If there is the possibility to  combine 
this incomplete 
  information about $\alpha^2F(\omega)$  with the 
results of other spectral measurements, for example, tunnelling or neutron 
data, one can  qualify the transport spectral function.
The so called ''model 2`` electron-boson interaction  function \cite{shulga91}
 was recovered assuming, that
the solution  has to be resemble the phonon density of states, measured 
by inelasitc neutron spectroscopy.

\subsection{Optical mass and impurity scattering rate}
\label{OmasGam}

In ref.\cite{shulga91,jos1} we shown, that it is  useful to describe
 the
 optical conductivity in terms of the so called ``extended''
Drude formula
\begin{equation}
\sigma(\omega,T)=\!\frac{\omega^2_{pl}/4\pi}{ W(\omega,T)}=
\frac{\omega^2_p/4\pi}{
\gamma_{opt}(\omega,T)-i\omega m^*_{opt}(\omega,T)/m_b}.
\label{eqw}
\end{equation}
Here the optical relaxation $W(\omega)$,
the optical mass $m^*_{opt}(\omega,T)/m_b$ and the
optical scattering rate $\gamma_{opt}(\omega,T)$ as the complex, real and imaginary
parts of an inverse normalised conductivity $\omega^2_{pl}/4\pi\sigma(\omega,T)$
have been introduced.
This presentation is old and popular in spectroscopy, but the practical value of the
Eq.\ (\ref{eqw})
has not been exploited much.
The optical relaxation will be compared with the
 effective mass $m^*_{eff}(\omega,T)/m_b$ 
and the effective scattering rate $\gamma_{eff}(\omega,T)$, which by
definition are  the real and imaginary
parts of the renormalized frequency
 \begin{equation}
\tilde{\omega}(\omega,T)=\frac{\omega m^*_{eff}(\omega,T)}{m_b}+
\frac{i\gamma_{eff}(\omega,T)}{2}.
\label{eqo}
\end{equation}

\begin{figure}[t]
\centerline{\hbox{   
\hspace{-2.3cm} 
\psfig{figure=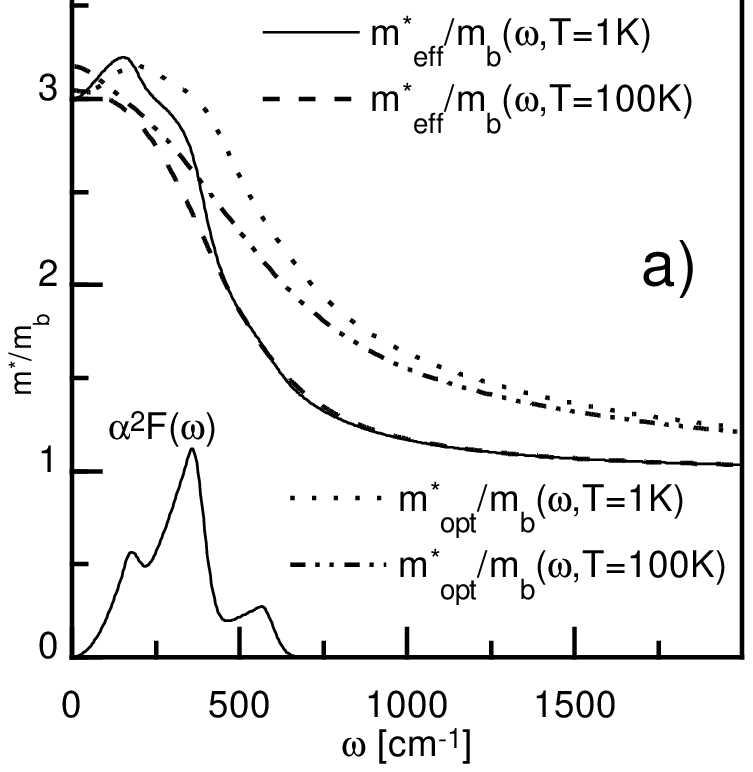,width=6.0cm,height=6.0cm}
\hspace{0.8cm} 
\psfig{figure=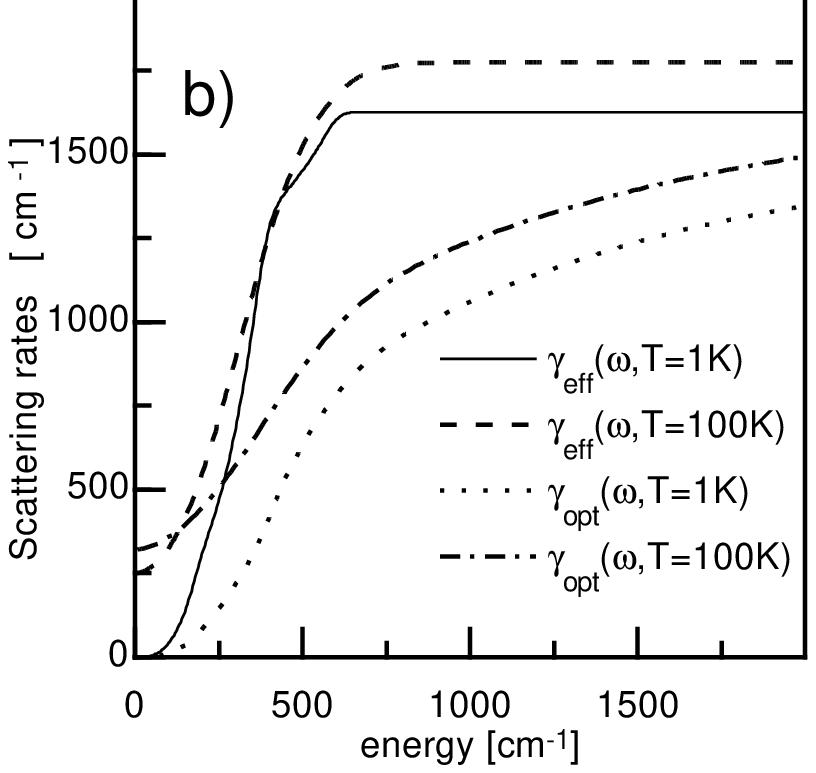,width=6.0cm,height=6.0cm}
}}
\vspace{1cm}
\caption{The frequency dependencies of the normal state optical and 
effective masses (a) and scattering rates (b) calculated using the  ''model 2`` 
spectral function with $\lambda$=2 (shown in panel (a) for comparison).
The quasiparticle effective mass $m^*_{eff}(\omega)/m_b$ and scattering rate 
 $\gamma_{eff}(\omega)$ at $T=$1 K (solid lines, panels (a) and (b)) reproduce 
the fine structure of the spectral function, but at $T=$100 K (dashed lines)
this function are smeared out  due to averaging over  Fermi distribution.
At $\omega>\Omega_{max}$ the quasiparticles become ''undressed``, \ 
$m^*_{eff}(\omega)/m_b\approx 1$ and $\gamma_{eff}(\omega)\approx const $. 
 The optical mass $m^*_{opt}(\omega)/m_b$ and the scattering rate 
 $\gamma_{opt}(\omega)$ even at $T=$1 K (dotted lines) do not reproduce 
the structure of the spectral function due to the energy averaging,
 all the more at high temperature
ones ($T=$100 K, dashed-dotted lines). } 
 \label{fig1}
 \end{figure}

 When the frequency dependencies
of $\gamma_{opt}(\omega,T)$ and  $m^*(\omega,T)/m_b$ are plotted,
   the coupling constant, the average
boson energy and the upper boundary of spectral function are visually 
accessible (see Fig.\ \ref{fig1}).
The appropriate visual accessibility rules  can be derived   from the properties
 of the renormalized frequency 
 $\tilde{\omega}$ using the Allen formula Eq.\ (\ref{allg}) \cite{allen71}.   Since such
useful expression was originally obtained in the weak coupling 
approximation ($\lambda\ll 1$) only, below I rederive it for
the general case, including  strong coupling.

For any $k$ the values of $\lambda_k$ are positive and are of 
 the same order of magnitude, except may be $\lambda_0$.
It means, that in the zero order the  series $\tilde{\omega}_n$ defined by 
Eq.\ (\ref{ommats})
is an arithmetical progression. Consequently, according to the Gauss rule
for the arithmetical series the  denominators
$\tilde{\omega}_k+\tilde{\omega}_{n-k-1}$ in (\ref{signm}) at a given $n$
 do not depend on $k$, that is,  they   coincide.
In view of Eq.\ (\ref{eqw}), Eq.\ (\ref{signm}) takes the form
\begin{equation}
W_{n+1}\approx \tilde{\omega}_k+\tilde{\omega}_{n-k}.
\label{wapp}
\end{equation}
Hence we can expand  the  denominator  in (\ref{signm}) 
\begin{equation}
\frac{4\pi\sigma_n}{\omega_{pl}^2}\equiv\frac{1}{W_{n}}
=\frac{1}{n}\sum_{k=0}^{n-1}\frac{1}{W_{n}\left[ 1+
(\tilde{\omega}_k+\tilde{\omega}_{n-k-1}-W_{n})/W_{n}
\right]}
\end{equation}
to the first order in power of 
$(\tilde{\omega}_k+\tilde{\omega}_{n-k-1}-W_{n})/W_{n}$.
After some simple algebra we arrive at the  sought-for Allen formula
\begin{equation}
W_{n}=\frac{2}{n}\sum_{k=0}^{n-1}\tilde{\omega}_k.
\label{allg}
\end{equation}

\noindent

The real axis versions of Eq.\ (\ref{allg})  at $T=$0 look similar 
\begin{eqnarray}
W(\omega) & = &
\frac{1}{\omega}
\int_{0}^{\omega}dz 2\tilde{\omega}(z),
\label{maindT0}\\
\frac{\omega m^*_{opt}(\omega,T)}{m_b} & = &
\frac{1}{\omega}
\int_{0}^{\omega}dz \frac{2zm^*_{eff}(z,T)}{m_b},
\label{moptT0}\\
\gamma_{opt}(\omega) & = &
\frac{1}{\omega}
\int_{0}^{\omega}dz \gamma_{eff}(z).
\label{goptT0}
\end{eqnarray}
and   justify the following simple   description of 
 the interaction of light with  the conducting subsystem of the  
normal state metal. 
   When a photon with an energy $\omega$ 
has been absorbed,  {\it two} excited {\it virtual}
quasiparticles, an ''electron`` and a ''hole`` are created. 
 If  the first particle's   frequency is
 $\omega^{\prime}$, the second ones frequency  should be  $\omega-\omega^{\prime}$.
  The   exited ''electron`` and  ''hole`` 
relax to the Fermi level according to the  quasiparticle laws
 (see Eqs. \ref{nel},\ref{ommats}).
Since  $\omega^{\prime}$ can varies from 0 to   $\omega$,
the optical relaxation W($\omega$) is the  frequency-averaged
 ($\frac{1}{\omega}\int_0^{\omega}$), double (electron+hole $\rightarrow$\ factor 2) 
renormalized frequency $\tilde{\omega}$. 

  At $T=0$ the ISB effective scattering rate
 $\gamma_{eff}(\omega)\equiv2Im\tilde{\omega}(\omega)$
\begin{equation}
 \gamma_{eff}(\omega)  =   \gamma_{imp}+2 \pi \int_0^{\omega} dy
\alpha^2F(y)
\label{gamefT0}
\end{equation}
has a clear physical meaning.  The exited quasiparticles with the energy 
$\omega$  can relax by the emission of  virtual bosons with the energy varying 
the range from 0 to   $\omega$ or by the impurity scattering.

  At finite temperature the Allen formula
\begin{equation}
W(\omega)=
\frac{1}{\omega}
\int dz \frac{
\tanh{\left( \frac{\omega-z}{2T}\right) }
+\tanh{\left( \frac{z}{2T}\right) }}{2} 2\tilde{\omega}(z)
\label{maind}
\end{equation}
and the effective scattering rate   (\ref{nel}) have exactly 
 the same interpretation, since the only difference between them and 
Eqs.~(\ref{maindT0},\ref{maindT0},\ref{goptT0},\ref{gamefT0}) is the 
presence of the Fermi distributions instead of step functions
at $T$=0.  

Figure  \ref{fig1} shows the frequency dependencies of the normal state optical and 
effective masses and scattering rates calculated using the model 
spectral function \cite{shulga91} with $\lambda$=2.  
The results are plotted for the  low temperature
$T$=1 K and for  $T$=100 K which is close to $T_c$ and has the order of 
$\Omega_{boson}/2\pi T$, where    $\Omega_{boson}$=330 cm$^{-1}$  is the averaged 
boson frequency.    Both  $m^{*}_{eff}(\omega)/m_b$ and
  $\gamma_{eff}(\omega)$ show  the characteristic features corresponding 
the  peculiarities of the phonon spectrum.  When the frequency  $\omega$  
exceeds the upper energy  bound of the boson spectrum $\Omega_{max}$,
 the quasiparticle becomes
''undressed``. Its mass   $m^{*}_{eff}(\omega)/m_b\approx 1$ and 
effective scattering rate   $\gamma_{eff}(\omega)\approx const $, but of course,
the absolute value of $\gamma_{eff}(\omega\rightarrow\infty)$ could be huge.

  The optical mass
 $m^{*}_{opt}(\omega)/m_b$ and scattering rate
   $\gamma_{opt}(\omega)$ have no features at 
$\omega\approx\Omega_{max}$ due to the  energy averaging
(\ref{maindT0},\ref{moptT0},\ref{goptT0},\ref{maind}).  One can only {\it calculate}
the approximate value of $\Omega_{max}$  as the frequency where the 
  $m^{*}_{opt}(\omega)/m_b-1$ and  $\gamma_{opt}(\omega)$ reach  the halves
 of their maximum magnitudes.  Fig.\ \ref{fig1} shows,
 that this my rule can not be applied
even to the calculated frequency dependence of the 
optical scattering rate   $\gamma_{opt}(\omega)$,
 the more to the experimental curve, where {\it the high energy} behaviour is distorted
by the interband transition contribution to the dielectric
permeability.  In contrast the  frequency dependence of  the optical mass
  $m^{*}_{opt}(\omega)/m_b$ is suitable for the application of the simple
criterion
\begin{equation}
 m^*(\omega=\Omega_{max})/m_b-1\approx 0.5(m^*(\omega=0)/m_b-1).
\end{equation}
 Moreover its 
 {\it low energy} value $m^{*}_{opt}(\omega=0)/m_b\approx 1+\lambda_{tr}$  
 itself contains a useful piece of information.  Fig.\ \ref{fig1}a shows, that
 the approximate values of the
transport coupling constant $\lambda_{tr}\approx (m^{*}_{opt}(0)/m_b-1)$ and
the upper frequency bound of the electron-boson coupling function
$ [m^{*}_{opt}(\Omega_{max})/m_b-1 \approx 0.5(m^{*}_{opt}(0)/m_b-1)]$ 
are visually accessible.
  If the spectral function is wide
and its peak position (or averaged frequency) coincide roughly with the 
$\Omega_{max}$/2,   the optical relaxation W($\omega$) is smooth and 
the elucidation of $\Omega_{boson}$ at high temperature is impossible. 
 When an exotic 
superconductor with
the narrow, $\delta$-function like spectral  function will be discovered, 
 following this way 
one can easily construct  the recipe for the evaluation of the strength and 
the position of the single peak.

\subsection{My favourite polynom}
\label{favor}

\begin{figure}[t]
\centerline{\hbox{   
\hspace{-2.3cm} 
\psfig{figure=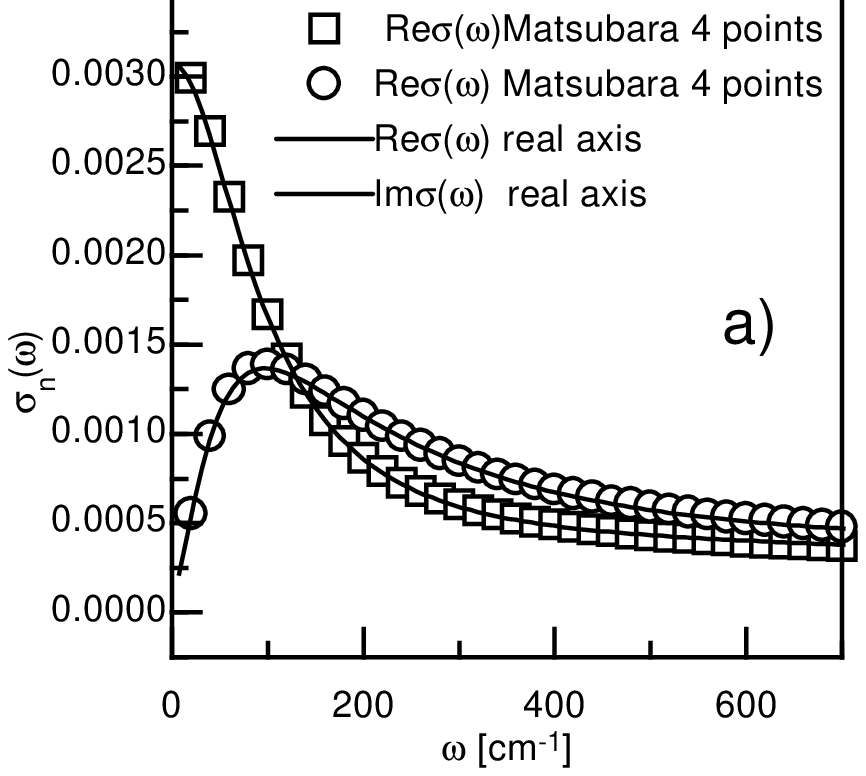,width=6.0cm,height=6.0cm}
\hspace{0.8cm} 
\psfig{figure=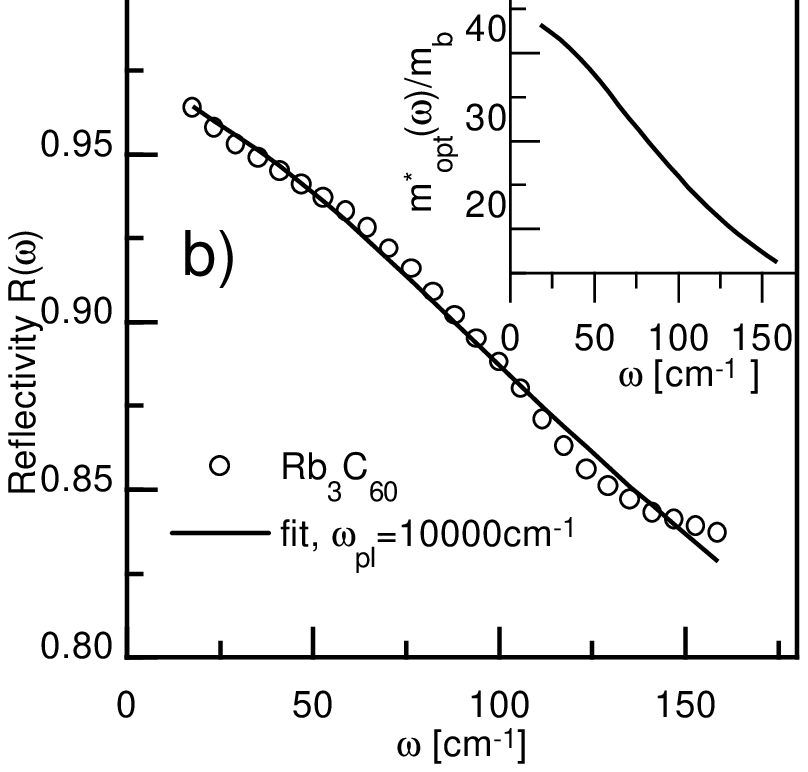,width=6.0cm,height=6.0cm}
}}
\vspace{1cm}
\caption{ Panel (a) shows the frequency dependencies of the real and imaginary parts
of the calculated normal state optical conductivity using the real  
axis formalism (solid lines)
and Matsubara technique (squares - Re$\sigma(\omega)$, circles - Im$\sigma(\omega)$).
The analytical continuation from the imaginary axis to the real one was made by the
Pade polynom with the degree of polynomial n=4. Calculations was made for T=100 K
using ''model 2`` spectral function.  Panel (b) shows the frequency dependence
of the normal state reflectivity R($\omega$) of Rb$_3$C$_{60}$ measured at T=40 K
by  Degiorgi {\it et al.,}
 {\it Phys. Rev. Lett.} {\bf 69}, 2987 (1992) (circles ) in comparison 
with the least square fit by the favourite polynom.
 Since four parameters were too much,
the value of the plasma frequency $\omega_{pl}$=10000 cm$^{-1}$ was fixed. 
The obtained frequency dependence of the  optical mass is shown in the inset.
  }
 \label{4p}
 \end{figure}

The Pade polynom analytical continuation \cite{Pade} is 
a capricious and not completely correct procedure. Nevertheless,
if the temperature is high enough in comparison with the energy
under consideration it works  splendidly.    The numerical experiment 
has given to my great surprise  the following (and to some extent obvious) 
 result (see Fig.\ \ref{4p}). The good agreement  between the real axis
calculations and the Matsubara ones takes place every time when 
the approximation using 1000 $\sigma_n$ points gives the same result as the {\it four} points 
Pade approximation.

 This remarkable Pade  polynom with  only four 
parameters $\omega_{pl}^2/4 \pi$,
$A$, $B$ and $C$
\begin{equation}
\sigma(\omega)=\frac{\omega_{pl}^2/4 \pi}{A
-i\omega+\frac{B\omega}{\omega+iC}}
\label{padx}
\end{equation}
fits well the majority of the experimental data at   $T\ge T_c$
and itself is the   demonstration  of the result obtained  in 
section \ref{exact}:  the  few
first Matsubara values $\sigma_k$  are responsible for the frequency
dependence of $\sigma(\omega)$.

The Pade polynom (\ref{padx}) is a formula of merit. At first, it has the correct 
analytical  properties. Its two  poles are located in the lower
half-plane of the complex energy.  The high  ($\omega\rightarrow\infty$ )
and low  ($\omega\rightarrow 0$) energy asymptotic behaviour are reasonable 
\cite{sulew88}.
 At second, it is  ideal for the 
least square fits of  experimental R($\omega$) \cite{burlakov92},
as well as for  simple analytical estimations.
Having the exact solution (see sec.\ref{exact})
 we  have to substitute  
 the complex values  $i\nu_k$ for energy, perform
easily the analytical continuation to the imaginary axis and analyse
the values of $\sigma_i$. In the way we obtain the interpretation of A, B, C. 
  Another possibility is 
a fast and easy analysis of the data in terms  
of  the optical mass $m^*_{opt}(\omega,T)/m_b$ and  the
optical scattering rate $\gamma_{opt}(\omega,T)$
\begin{eqnarray}
\frac{m^*_{opt}(\omega,T)}{m_b} -1& = & \frac{BC}
{\omega^2+C^2}
\label{4pmas} \\
\gamma_{opt}(\omega,T) & = &  A+\frac{B\omega^2}
{\omega^2+C^2}.
\label{4pgam}
\end{eqnarray}

At third, if the low-frequency reflectivity R($\omega$) measured at 
high temperature can not be fitted by Eq.\ (\ref{padx})  or the obtained value of
$\lambda_0$  contradicts  the generally accepted values,
it might indicate,  that the material under consideration is {\it not} a standard
Isotropic Single wide Band (ISB) metal and its properties
have to be analysed if terms of {\it another,} unconventional model. 
In order to illustrate the utility of the this approach, let us consider 
another superconducting compound with relatively high 
transition temperature $Rb_3C_{60}$ \cite{degeorgy92,mazin93,degeorgy94}.
Since the ratio of 
the  experimental reflectivities in the superconducting and normal states 
$R_s/R_n\approx 1$
above 100 cm$^{-1}$ \cite{degeorgy92}, in frame of the ISB model
it means, that the high energy bound of the electron-boson interaction
function $\Omega_{max}<$ 100 cm$^{-1}$.  
The low energy part
of the normal state reflectivity R($\omega$), measured at T=40 K,
was fitted (with the adopted fixed $\omega_{pl}$=10000 cm$^{-1}$)
by  (\ref{padx})  (see Fig. \ref{4p}b), $m^*_{opt}(\omega,T)/m_b$ is shown in the inset. 
 One can see, that the optical mass at small $\omega$ is huge itself and
 gives $\lambda > 100$. This  values points  to   heavy fermion 
or polaron models, rather than to the generally accepted weak or intermediate coupling
 scenario. 
On the other hand, the analysis of the normal state specific heat data
gives an extremely small value of the coupling constant.
 As a result we concluded that Rb$_3$C$_{60}$ is not 
a ISB metal and its properties should be treated in terms of a more
complicated model which takes into  account
the self-energy
 and
 conductivity vertex corrections, 
the small value of the bandwidth  $\Delta E\approx$100 meV,
possible influence of the electron-electron correlation, the strong
anisotropy of the coupling function and the Fermi velocity,  and etc. 
 
And the last but not least,  if the data can be fitted by  (\ref{padx}),
that is by the formula with  four parameters,  it is 
impossible to recover from this {\it four}  parameters  the detailed
 shape of the electron-boson
interaction function $\alpha^2F(\omega)$.

In conclusion, it is naturally to define the amount of the available
information  by  the number of poles and zeros of the function
which fits well the experimental curve or 
 the number of the
determinate Matsubara values.

\subsection{Temperature dependence of the optical mass and relaxation rate
 and its two-band  ersatz}

\begin{figure}
\centerline{\hbox{   
\hspace{-2.3cm} 
\psfig{figure=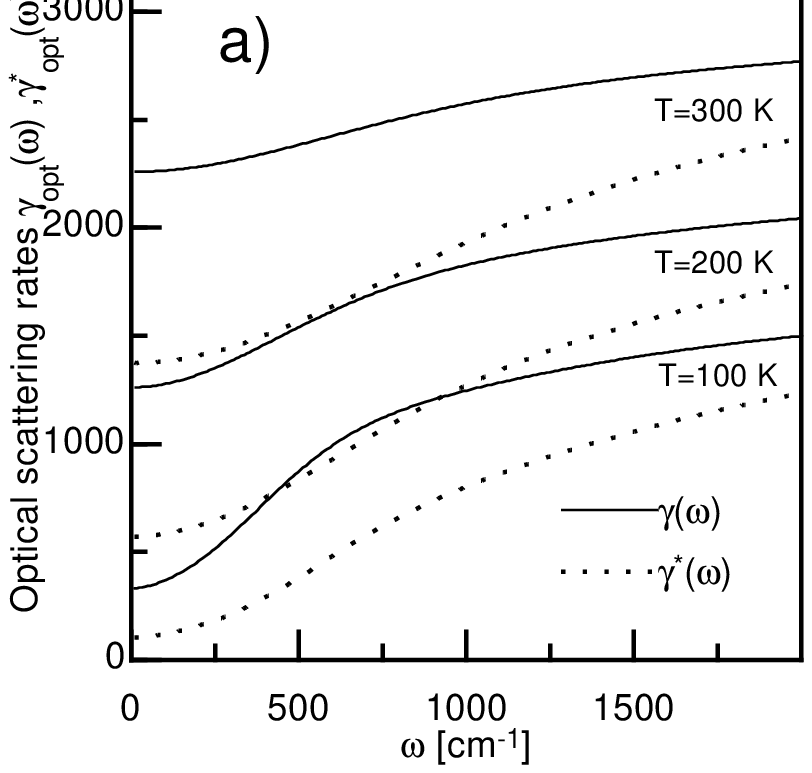,width=6.0cm,height=6.0cm}
\hspace{0.8cm} 
\psfig{figure=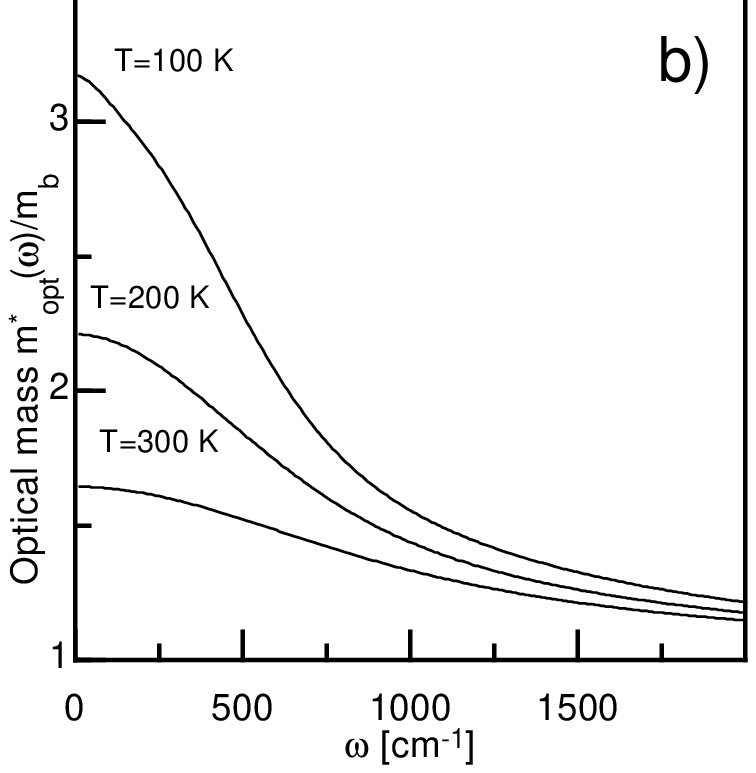,width=6.0cm,height=6.0cm}
}}
\vspace{0.7cm}
\centerline{\hbox{   
\hspace{-2.3cm} 
\psfig{figure=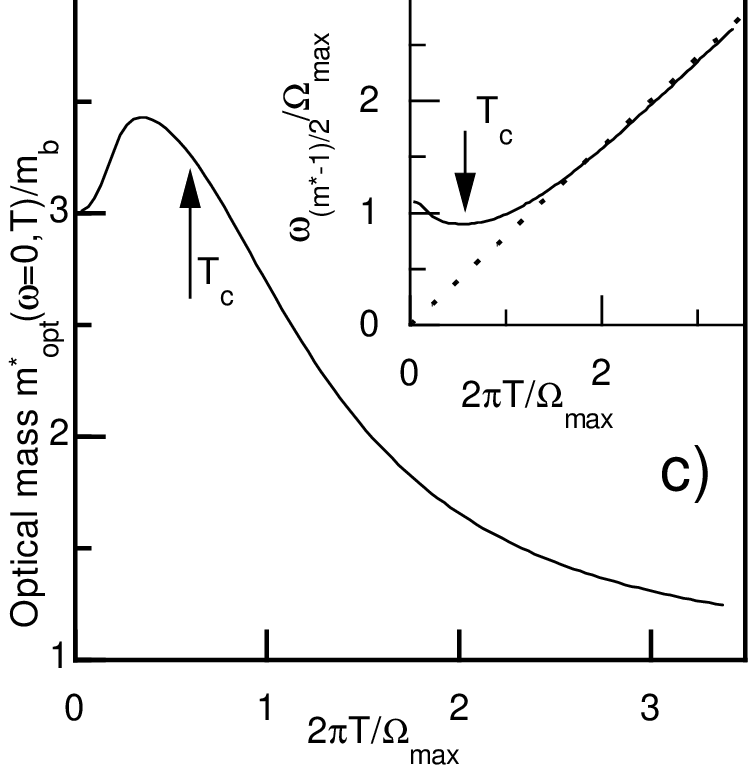,width=6.0cm,height=6.0cm}
\hspace{0.8cm} 
\psfig{figure=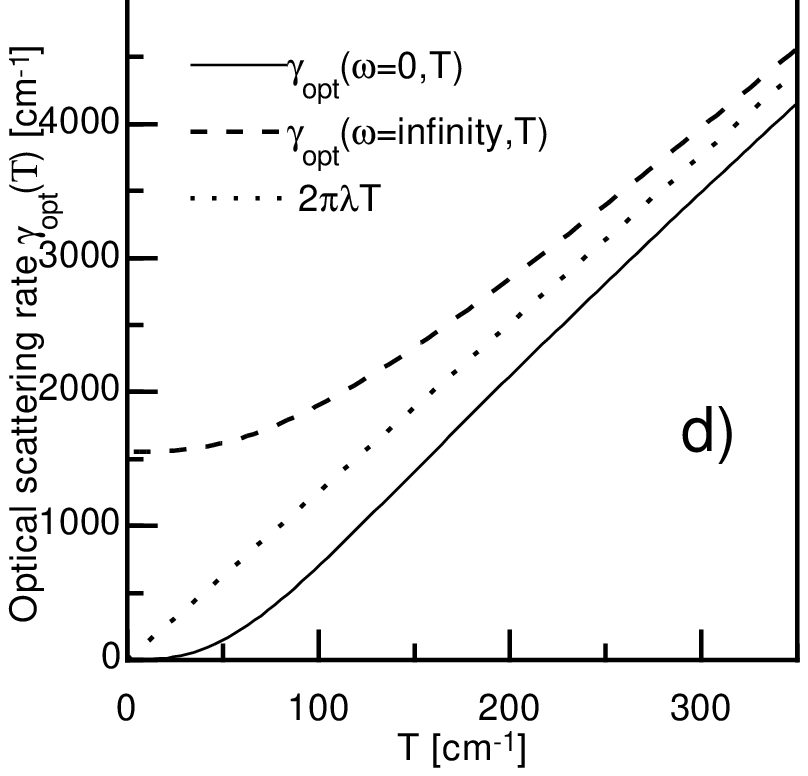,width=6.0cm,height=6.0cm}
}}
\vspace{1cm}
\caption{Panels (a) and (b) show  the frequency dependence 
of the optical mass 
and the optical scattering rate for T=100K, 200 K, 300 K.
Panels (c) and (d) present
 the corresponding  temperature dependence of $m^*_{opt}(T)/m_b$
and $\gamma_{opt}(T)$ for $\omega=0$ and  $\omega=\infty$.
All curves (solid lines) was calculated using model~2 spectral function.
For comparison the normalised optical scattering rates $\gamma^*(\omega,T)=
\gamma_{opt}(\omega,T)/m^*_{opt}(\omega,T)/m_b$
are plotted   by dotted lines in panel (a).
The temperature dependence of the quantity $\omega_{(m^*-1)/2}$ (see text) 
is shown by solid line on inset (c) together 
with the phenomenological line $\omega=5T$ (dotted line).
The  common for $\gamma_{opt}(\omega=0,T)$ and $\gamma_{opt}(\omega=\infty,T)$
asymptote  $\gamma=2\pi\lambda T$ is shown by dotted line in panel (d). }
 \label{gammasott}
 \end{figure}

In frame of the ISB model we assume, that all quasiparticles have the 
same Fermi velocity, but in real metals there is some anisotropy of $v_F$.
If the impurity scattering rate $\gamma_{imp}$ is small in comparison with the
complete optical scattering rate $\gamma_{opt}$  and the electron-boson coupling
is isotropic,  
we can use the average value $<v_F^2>$ and keep the ISB model.
On the other hand, if the impurity scattering, given by the mean free path $l$,
dominates, the quasiparticles from  different parts of the Fermi surface
will have different scattering rates $\gamma_{imp,i}=v_{F,i}/l$. Let us  accept for
simplicity, that we can divide the Fermi surface between two parts 
with the approximately uniform properties. The optical conductivity of this two-band system
is the sum of two Drude terms
\begin{equation}
\sigma(\omega)=\frac{\omega_{pl,1}^2/4\pi}{\gamma_1-i\omega}
+\frac{\omega_{pl,2}^2/4\pi}{\gamma_2-i\omega}.
\label{2drude}
\end{equation}
One can see, that Eq.\ (\ref{2drude}) and Eq.\ (\ref{padx}) are equivalent and four parameters 
$\omega_{pl,1}^2$,$\omega_{pl,2}^2$, $\gamma_{1}$, $\gamma_{2}$ in 
(\ref{2drude}) connected with $\omega_{pl}^2$, $A$, $B$, $C$
 in (\ref{padx}) as follows
\begin{eqnarray}
\omega_{pl}^2&  = &\omega_{pl,1}^2+\omega_{pl,2}^2 \nonumber \\
C & =& \frac{\omega_{pl,1}^2\gamma_2+\omega_{pl,2}^2\gamma_1}
{\omega_{pl,1}^2+\omega_{pl,2}^2} \nonumber \\
A & =& \frac{\gamma_1\gamma_2(\omega_{pl,1}^2+\omega_{pl,2}^2)}
{\omega_{pl,1}^2\gamma_2+\omega_{pl,2}^2\gamma_1}
 \nonumber \\
B & =& \gamma_1+\gamma_2-A-C
\end{eqnarray}
It means, that  even if there is no electron-boson interaction
in the two-band system, 
the frequency dependence of its optical 
conductivity  (\ref{2drude})  will reproduce a ISB model one (\ref{signr}) with some 
frequency dependent ''mass`` and ''scattering rate``.  The multiband ersatz of the
electron-boson effects could be revealed by the investigation of the 
temperature dependence of  $m^*_{opt}(\omega,T)/m_b$ and $\gamma_{opt}(\omega,T)$.
It originates mainly from 
the  presence of the Bose factor in Eqs.(\ref{nel}) ($\equiv$ sampling in (\ref{ommats})) 
and to a less degree from the Fermi distributions in Eqs.\ (\ref{nel}-\ref{signm}). 

In subsection \ref{OmasGam} the low  ($T\rightarrow 0$) 
 temperature behaviour 
of  $m^*_{opt}(\omega)/m_b$
and $\gamma_{opt}(\omega)$ was discussed in detail. We found, that
at a given quasipartical frequency $\omega$ in accord with the
frequency conservation principle only
 virtual bosons 
with $\Omega<\omega$ can contribute to the relaxation.

At high 
temperature the  bosons lose their individuality. 
At $T\rightarrow\infty$ the expression (\ref{nel}) can be substantially simplified, since 
the Bose factor $coth(z/2T)\approx 2T/z$
 dominates in comparison with the Fermi one  $tanh[(\omega-z)/2T]$. Keeping 
the leading term $2T/z$ we arrive at
\begin{equation}
Im\tilde{\omega}(\omega)\equiv\gamma_{eff}(\omega)\approx \frac{\gamma_{imp}}{2}+
\pi T\int\frac{dy\alpha^2F(y)}{y}= \frac{\gamma_{imp}+2\pi\lambda T}{2}.
\label{ratb}
\end{equation}
 In Matsubara technique the same result can be obtained even more easily.
 At $T\rightarrow\infty$  
all kernels of the spectral function 
$\lambda_{k}$,  defined by Eq.\ (\ref{aaa2}),  vanish as
 $1/(<\Omega^2_{boson}>+4\pi^2 k^2 T^2)$
 for all k, except k=0. Therefore,
the Eq.\ (\ref{ommats}) for the renormalized frequency becomes trivial 
\begin{equation}
\tilde{\omega}_n=\omega_n+\gamma_{imp}/2+\pi\lambda T.
\label{matstb}
\end{equation}
If one substitutes the obtained $\tilde{\omega}(\omega)$ in Eq.\ (\ref{signr})
or $\tilde{\omega}(i\omega_n)$ into  Eq.\ (\ref{signm}) after some algebra
one arrives at the Drude type expression for the optical conductivity
\begin{equation}
\sigma(\omega)=\frac{\omega^2_{pl}}{4\pi}\frac{1}
{\gamma_{imp}+2\pi\lambda T-i\omega}.
\label{condt}
\end{equation}

Note, that the relation $\gamma_{opt}=2\gamma_{eff}\equiv -2Im\Sigma$ is correct,
if and only if the self energy $\Sigma(\omega)=\omega-\tilde{\omega}(\omega)$
does not depend on frequency. Otherwise, 
the relation 
\begin{equation}
\gamma_{opt}(\omega)=-2Im\Sigma(\omega)
\label{varma}
\end{equation}
does not satisfy the
Kramers-Kronig (K.-K.) relations, that is, it violates the causality.
 The proof is very simple.
Due to the charge conservation law, 
when the photon has been absorbed,
{\it two} quasiparticles, ''the electron``
and ''the hole'' come into being. The optical 
conductivity is a bosonic function defined by its values at $\nu_k$, while
the quasiparticle self energy $\Sigma(\omega)$ is a fermionic function
defined by the values at $\omega_k$. 
As a result, the right and left  hand sides of Eq.\ (\ref{varma})
  have different statistics,
what is very strange. In frame of the ISB model considered here 
 the proof is a little bit more complicated. 
The optical conductivity itself and its inverse
$1/\sigma(\omega)$ are  response functions. It  means, that 
they have no poles on  the  upper half-plane of the
complex frequency or, in other words, their real and imaginary
parts are connected by Kramers-Kronig relations.
 The same is valid for  the self energy. 
 From Eq.\ (\ref{varma}) after K.-K. transformation  one obtains
 \begin{equation}
m^{*}_{opt}(\omega)-1=2(m^*_{eff}(\omega)-1).
\label{varmam}
\end{equation}
The optical and effective masses have the same asymptotic value 1+$\lambda$
 at $\omega,T\rightarrow 0$. 
Due to the presence of the factor of 2 in (\ref{varmam}) it would mean, that 1=2.  
  
\begin{figure}
\centerline{\hbox{   
\hspace{-2.3cm} 
\psfig{figure=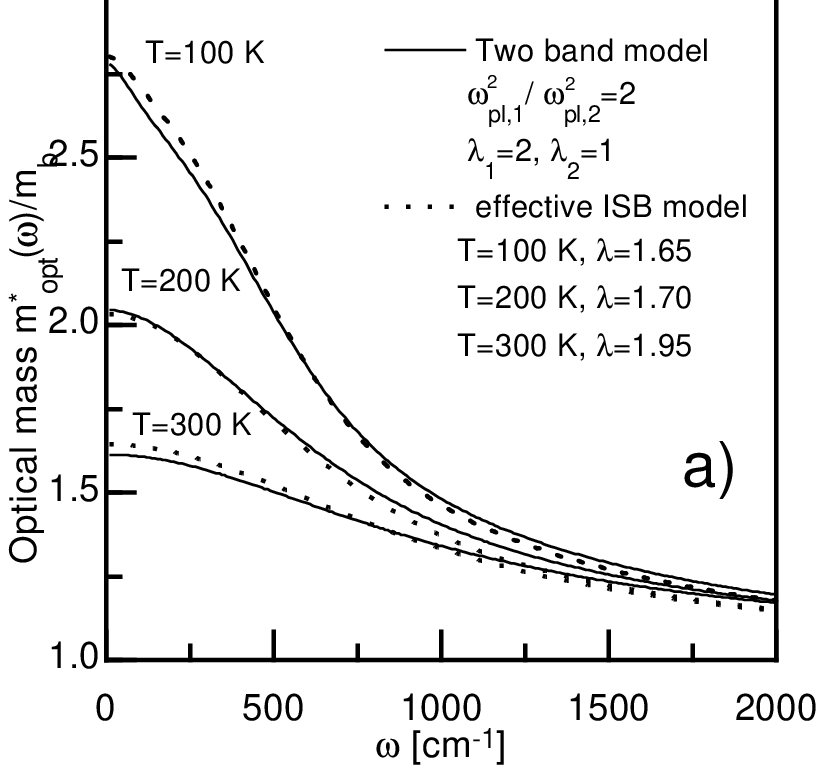,width=6.0cm,height=6.0cm}
\hspace{0.8cm} 
\psfig{figure=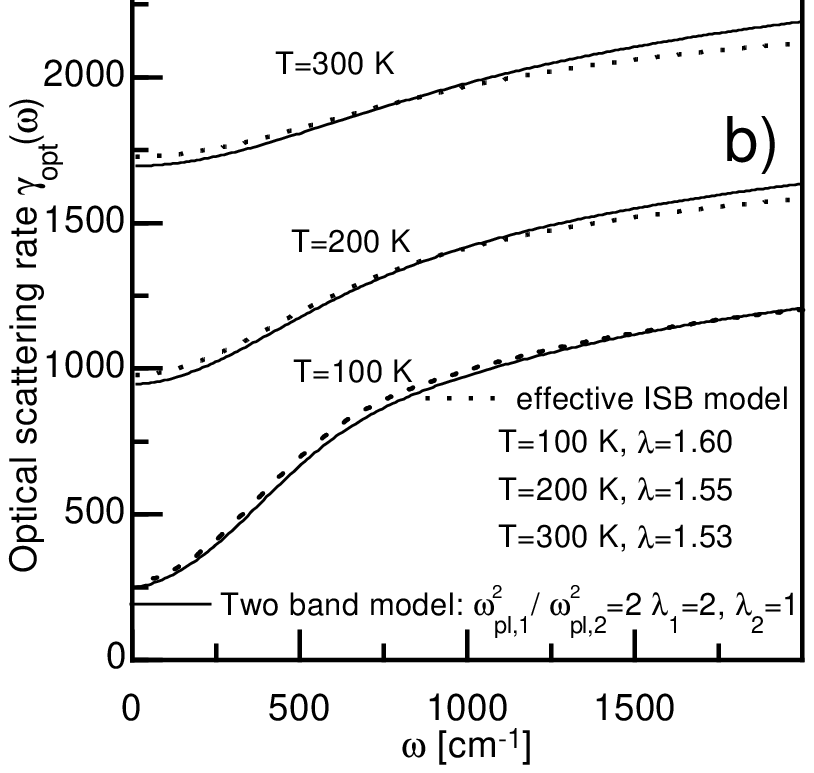,width=6.0cm,height=6.0cm}
}}
\vspace{0.7cm}
\centerline{\hbox{   
\hspace{-2.3cm} 
\psfig{figure=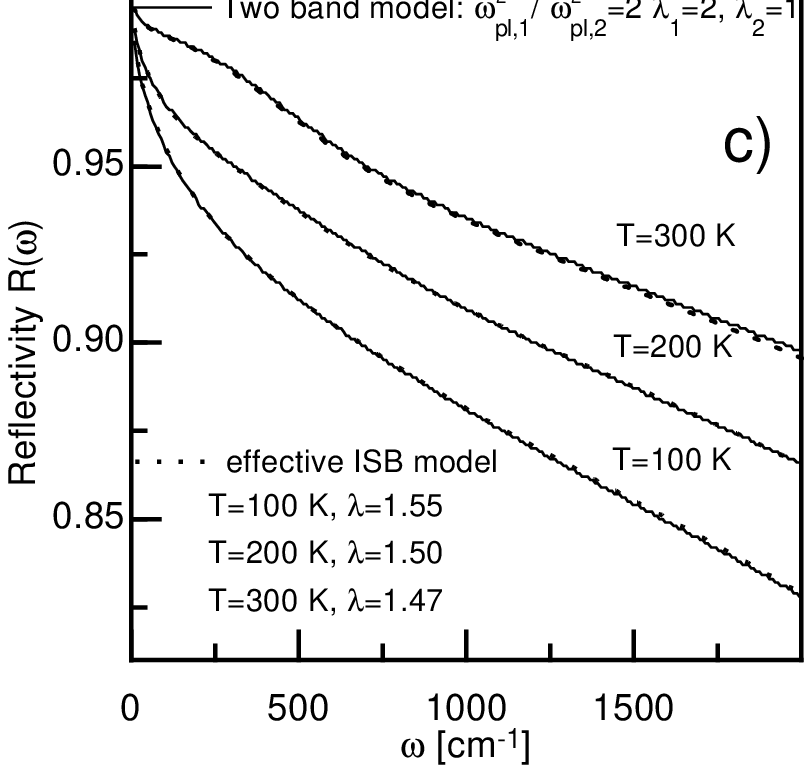,width=6.0cm,height=6.0cm}
\hspace{0.8cm} 
\psfig{figure=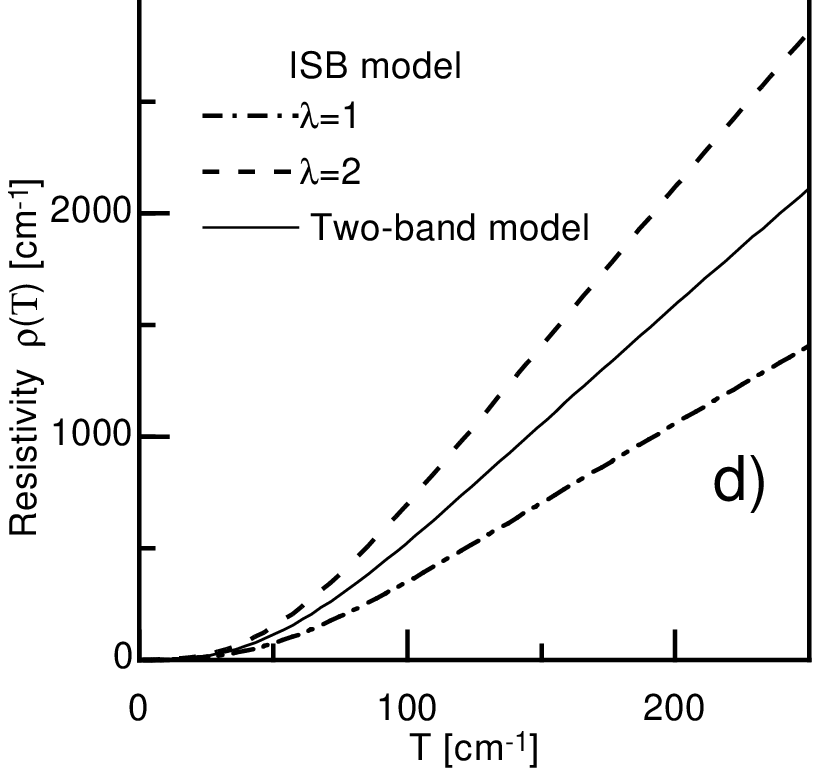,width=6.0cm,height=6.0cm}
}}
\vspace{1cm}
\caption{The frequency and temperature dependencies of the two-band 
optical mass, scattering rate, reflectivity and DC resistivity 
are shown by solid lines  in panels a,\ b,\ c,\ d respectively.
The corresponding ISB fitted counterparts are plotted by the dotted lines. }
 \label{twoband}
 \end{figure}
Let us turn to the temperature dependence of the optical properties 
of ISB metals, which are  shown in Fig.\ \ref{gammasott}. 
All curves were calculated using model~2 spectral function and $\gamma_{imp}=0$.
Panels \ref{gammasott}a presents the frequency dependence of the 
optical scattering rate $\gamma_{opt}(\omega,T)$ (solid lines) for 
T=100 K, 200 K, 300K. The corresponding optical masses are 
shown in panel  \ref{gammasott}b. One can see, that the absolute value 
of   $\gamma_{opt}(\omega,T)$  increases with T, but the frequency dependent
component decreases. It leads to diminishing of  $m^*_{opt}(\omega,T)/m_b$ 
when T is rising. The temperature dependence of the 
minimum ($\omega=0$,solid line) and maximum ($\omega=\infty$,solid line) 
values of $\gamma_{opt}(\omega)$  are shown in panel \ref{gammasott}d
in comparison with  
their common asymptote $\gamma=2\pi\lambda T$ (dotted line). The corresponding 
function  $m^*_{opt}(\omega=0,T)/m_b$, named sometimes $1+\lambda_{opt}(T)$ is
presented in panel 5c. The position of maximum of the small hump at low  T
is approximately equal to $\Omega_{min}/5$, where $\Omega_{min}$ is the
frequency position of the lowest boson peak in $\alpha^2F(\omega)$. At the same
temperature the DC resistivity $\rho(T)=4\pi\gamma_{opt}(\omega=0,T)/\omega_{pl2}^2$ 
starts the linear rise. In subsection  \ref{OmasGam} the simple rule for determination
of the {\it upper} frequency bound of the spectral function $\Omega_{max}$ was proposed.
At $\omega\approx\Omega_{max}$  the optical mass without unity 
$(m^*_{opt}(\Omega_{max},T)/m_b-1)$ is approximately  the half of its value at 
$\omega=0$.  The temperature dependence of the quantity $\omega_{(m^*-1)/2}$ defined by
\begin{equation}
2[m^*_{opt}(\omega=\omega_{(m^*-1)/2},T)/m_b-1)]=
m^*_{opt}(\omega=0,T)/m_b-1
\end{equation} 
is shown by solid line on inset in panel \ref{gammasott}c in comparison
with the phenomenological line $\omega=5T$. One can see, that 
\begin{equation}
\omega_{(m^*-1)/2} \approx max(5T,\Omega_{max})
\end{equation}
The normalised optical scattering $\gamma^*(\omega)=
\gamma_{opt}(\omega)/m^*_{opt}(\omega,T)/m_b$
is shown in Fig(\ref{gammasott}a) (dotted lines). In our opinion \cite{shulga91}
this quantity  is far useless in comparison with  $\gamma_{opt}(\omega)$ and 
$m^*_{opt}(\omega,T)/m_b$. 
With the exception of the low frequency region where its value is  constant,  
$\gamma^*(\omega)$ is 
a  quasilinear  function  $\propto\omega$ up to max$(4\Omega_{max},2\pi\lambda T)$.
Even the simple analysis by the favourite polynom Eq.\ (\ref{padx}) allows to determine the
four parameters, while $\gamma^*(\omega)$ depends on two. It is interesting, that 
the Gilbert transformation from  $\gamma^*(\omega)$  is not equal
to 0. In this connection we note that the use of the BSC normal state conductivity expression
\begin{equation}
\sigma(\omega)=\frac{\omega_{pl}^{*2}}{4\pi(\gamma^*-i\omega)},
\end{equation}
is correct, only if $\gamma^*=const$.
Here  $\omega_{pl}^{*2}=\omega_{pl}^{2}/(1+\lambda)$
is the frequency-independent renormalised plasma frequency.

Note, that according to the Allen formula (\ref{maind}), the optical relaxation
is approximately {\it linear} with respect to the spectral function $\alpha^2F(\omega)$.
It means, that the conversion from the low temperature scenario to the asymptotic
high temperature one takes place for each boson with frequency $\Omega_i$
separately at $T \approx \Omega_i/5$, that is, when   $\Omega_i$
becomes of the order of the {\it first} Matsubara energy $\nu_1=2\pi T$. 

Since the isotropic single wide band  model is the subject of the present lecture,
 discussing the properties of 
two-band systems  we have to pay special  attention to
possible artefact results. Let us start our consideration from 
the temperature independent two-Drude terms simulator (\ref{2drude}).
Since it is equivalent to the favourite polynom (\ref{padx}), we can ''extract
the information about spectral function``  including the temperature 
dependence of the coupling constant $\lambda(T)$.  It is known, that if 
the material parameters are temperature independent, the optical scattering rate
$\gamma_{opt}(\omega,T)$ is a monotonically increasing function.
Since the simulator's $\gamma_{opt}(\omega)$ does not changes, when T is rising,
one may  conclude,  that the coupling ''constant`` $\lambda(T)$ is a 
monotonically decreasing function  with the high temperature asymptote 1/T.
On the other hand,  the analysis of the optical  mass, which is usually  a
monotonically decreasing function, gives the opposite result: $\lambda(T)$ dramatically
increases when temperature is growing. The study of the reflectivity $R(\omega,T)$
 gives the third
fitted function  $\lambda(T)$.  The reasonable way to resolve this conflict is the
replacement of the model.  If one restricts oneself to the consideration of the scattering rate
only, the contradiction can be overlooked. The fact is that the strong temperature
dependence of material parameters exists, for example, in heavy fermion systems.
Another interesting scenario was realised in  fullerides, where
the DC resistivity (measured at the constant pressure) changes with temperature as T$^2$
and the same DC resistivity (measured at the constant volume) follows conventional linear
law $\rho\propto T$, as it should be, since the theory assumes the V=const condition.
 
 The second example  is the clean limit of anisotropic systems.  As usually,
one  can divide the Fermi surface between two (for simplicity)  parts 
with the approximately uniform properties.  This disjoined representation of the 
Fermi surface is known as the multiband model.  We accept also, that the intra- 
and interband interactions are governed by the electron-boson
spectral functions having the same ''model 2`` shape, but different coupling
constants $\lambda_{ij}$.  The scales of  transport and
optical properties of the bands are fixed by the plasma frequencies
  $\hbar \omega_{pl,i}^2=4\pi e^2 v_{F,i}^2 N_i(0)/3$. 
In the normal state is possible to diagonalise the equations by the substitution
$\lambda_1=\lambda_{11}+\lambda_{12}$ and 
$\lambda_2=\lambda_{21}+\lambda_{22}$. The resulting
two-band optical conductivity is the sum of two ISB terms (\ref{signr}).  
For the model calculations 
the arbitrary values $\lambda_1=2$, $\lambda_2=1$ and
 $\omega_{pl,1}^2/ \omega_{pl,2}^2$=2 was chosen. 

The frequency and temperature dependencies of the two-band 
optical mass, scattering rate, reflectivity and DC resistivity 
are shown by solid lines  in Fig.\ (\ref{twoband})a,\ b,\ c,\ d respectively. 
 The {\it by-eye-analysis}
can not reveal the difference between clean limit two-band optical properties and
the standard ISB ones.  Fortunately, the ISB model can pose several 
possibilities for a quantitative study. I simply  fit the two-band  curves, shown on
panels (\ref{twoband})a,\ b,\ c,\ 
by the appropriate ISB ones (dotted lines),
calculated using the same ''model 2`` spectral function. 
The coupling constant $\lambda$ was used as fitting parameter.
 One can see, that  the temperature dependencies of $m^*_{opt}(\omega,T)/m_b$,
 $\gamma_{opt}(\omega,T)$ and $R(\omega,T)$
of the anisotropic  system are yet weaker, that it should be according 
to the ISB rules. The fitted functions $\lambda(T)$ grow or drop, when T is
rising, depending on which quantity has been chosen for the fit.

For the few  metals, where the mutual anisotropy
of $\lambda({\bf k})$ and of $v_F({\bf k})$ was investigated, the following 
regularity was found. The quasiparticles with smaller  $v_F({\bf k})$ 
have larger  $\lambda({\bf k})$.  Usually this information is unavailable.
 One can estimate 
the anisotropy, if one compares the material parameters obtained from 
the optical spectra with the ones revealed by the analysis of 
 thermodynamic, or point-contact tunnelling data.
The scale of the 
optical conductivity is fixed by  $\omega_{pl}^2\propto v_F^2N(0)$. 
The normal state specific heat data 
are proportional to $(1+\lambda)N(0)$. The point-contact tunnelling spectra are
weighted by the factor $v_{F}$. 
As a result, if one  treats  optical spectra of 
anisotropic metals 
in terms of the ISB the model, one 
reveals the electron-boson coupling function $\alpha^2_{opt}F(\omega)$,
which distinguish itself by smaller value of the coupling function $\lambda_{opt}$
in comparison with lambdas, obtained by other physical methods. 
This difference is  usually assigned to the kinetic coefficient
 (1-cos) \cite{allen71} or,
in another language, to the vertex corrections to the quasipartical 
 Green function\cite{nam67}. 
It is interesting, that if the system under consideration is completely isotropic,
and $\alpha^2F(\omega,{\bf k}, {\bf k}^{\prime})$ do not depend on {\bf k} and
 {\bf k}$^{\prime}$, the vertex corrections vanish as soon as considered in this 
subsection usual anisotropy effects. What kind of corrections are more
important in each particular case, first order vertex ones or discussed 
 here zero order distortions caused by 
the interplay of $v_f(\bf k)$ and $\lambda({\bf k})$, it is the question
{\it for} the gallop band structure calculations.  
 
 In conclusion, the temperature dependence of the optical properties 
of ISB metals really exists and can be well quantitatively described. 
It originates mainly from 
the  presence of the Bose factor in Eqs.(\ref{nel}) ($\equiv$ sampling in (\ref{ommats})) 
and to a lesser extent from the presence of the
Fermi distributions in Eqs.(\ref{nel}-\ref{signm}).  The principal deviation from 
the predicted frequency and temperature behaviour manifests 
that the chosen ISB model is not complete and has to be replaced. 

\subsection{Optical mass and scattering rate in mid-infrared: artefacts generated by
overlooking of interband transitions and underestimation of $\epsilon_{\infty}$.}

The formal textbook relation between the conductivity $\sigma(\omega)$  and 
permeability $\epsilon(\omega)$  
\begin{equation}
\epsilon(\omega)=1+\frac{4\pi i \sigma(\omega)}{\omega}
\label{idiotsigma}
\end{equation}
is not suitable for a practical use, since the entering  Eq.\ (\ref{idiotsigma})
$\sigma(\omega)$  reflects the total response  of all excitations. 

The simplest reasonable decomposition of (\ref{idiotsigma}),
 valid over the spectral region up to  frequencies just above plasma edge, is 
\begin{equation}
\epsilon(\omega)=\epsilon_{\infty}+\epsilon_{inter}(\omega)+\epsilon_{phonon}(\omega)
+\frac{4\pi i \sigma(\omega)}{\omega},
\label{noreps}
\end{equation}
where $\epsilon_{inter}$ and $\epsilon_{phonon}$ are the
contributions of interband transition(s) and  direct IR active phonons
centred below plasma edge.  The response of high energy excitations and
 wide interbands, overlapping the plasma edge is presented by the complex
constant   $\epsilon_{\infty}$. $\sigma(\omega)$ is {\it the quasiparticle }
conductivity, defined by Eqs.(\ref{signr},\ref{sigsup}). 

The direct phonon band contribution will not be considered,
since it is usually weak and plays some role only  in 
the selected compounds containing 
light elements and having small value of $\omega_{pl}$, for example,
in doped fullerenes.  

The used here model interband contribution 
\begin{equation}
\epsilon_{inter}(\omega)=\frac{\omega_{pl,inter}^2}
{\Omega_0^2-\omega^2-i\omega\gamma_{inter}}
\label{epsinter}
\end{equation}
is the simple  Lorentzian with $\omega_{pl,inter}$=16000 cm$^{-1}$,
$\Omega_0$=4000 cm$^{-1}$, and 
$\gamma_{inter}$=6000 cm$^{-1}$. 
The overlooking of the interband contribution was simulated as follows.
The sum of two terms given by Eqs. (\ref{epsinter}) and (\ref{signr})
was treated as the quasiparticle response in terms of the optical 
mass and scattering rate
\begin{equation}
\gamma_{opt}(\omega,T)-i\omega m^*_{opt}(\omega,T)/m_b
=\frac{\omega_{pl}^2+\omega_{pl,inter}^2}
{4\pi \sigma(\omega)-i\omega \epsilon_{inter}(\omega)}.
\label{epsinter2}
\end{equation}
For simplicity I set $\omega_{pl}=\omega_{pl,inter}$ and following to  

\begin{figure}
\centerline{\hbox{   
\hspace{-2.3cm} 
\psfig{figure=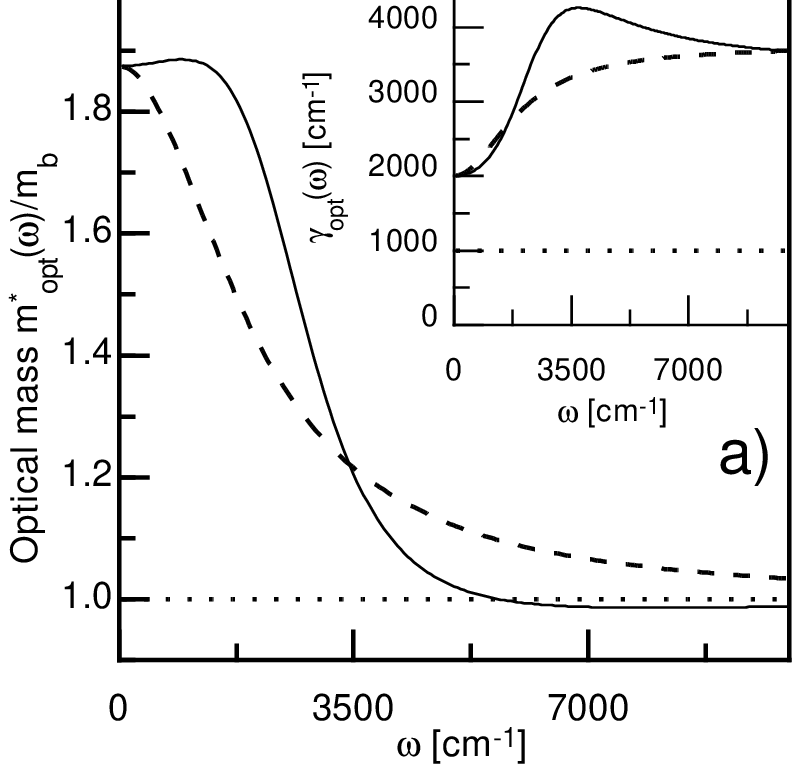,width=6.0cm,height=6.0cm}
\hspace{0.66cm} 
\psfig{figure=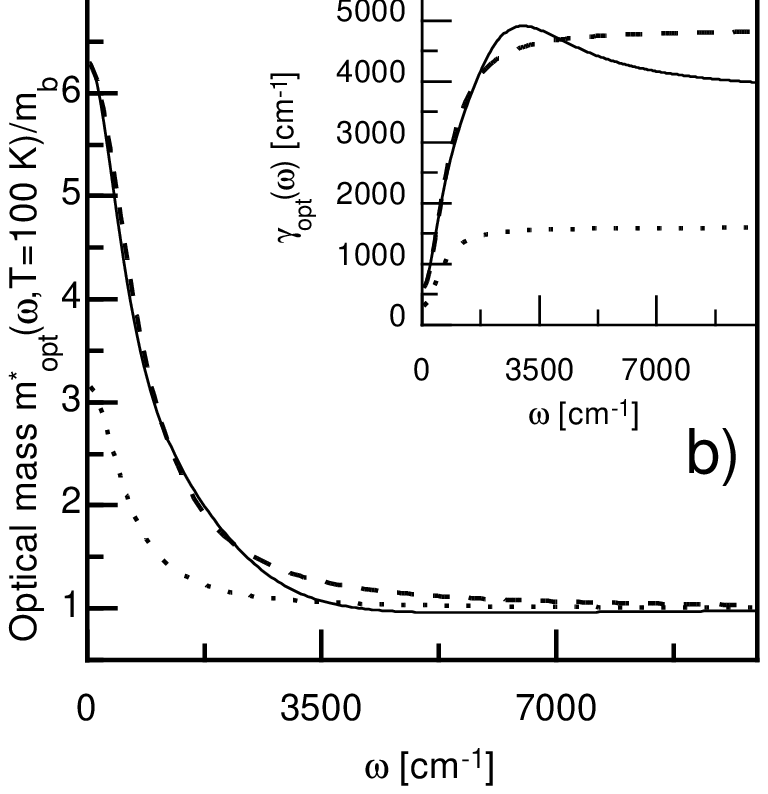,width=6.0cm,height=6.0cm}
}}
\vspace{0.7cm}
\centerline{\hbox{   
\hspace{-2.3cm} 
\psfig{figure=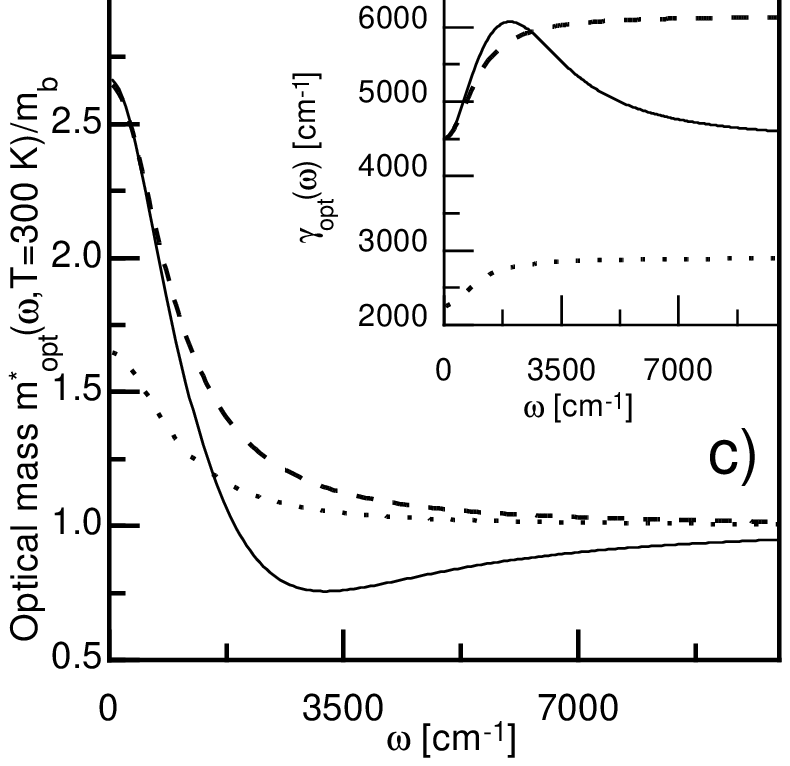,width=6.0cm,height=6.0cm}
\hspace{0.66cm} 
\psfig{figure=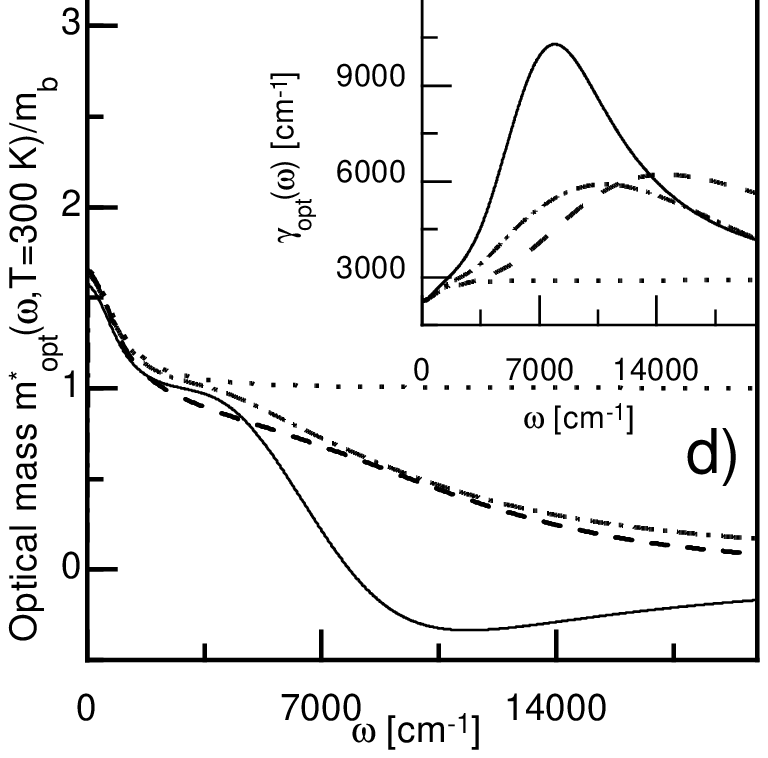,width=6.0cm,height=6.0cm}
}}
\vspace{1cm}
\caption{The artefact frequency dependencies  of 
the optical masses and scattering rates (insets)  generated by the 
overlooking of interband transition        (panels a,\ b,\ c) and the 
underestimation of $\epsilon_{\infty}$ (panel d) are shown by 
solid lines. The true, unpertubed  frequency dependencies  of 
$m^*_{opt}(\omega)/m_b$  and $\gamma_{opt}(\omega)$ are plotted by dotted lines.
The dashed lines in panels a,\ b,\ c are the {\it chi-by-eye} fitted curves ( by Eq.\ 32).
For parameters of the model interband transition and trial ISB quasiparticle
optical conductivities see text.} 
 \label{inter}
 \end{figure}

The results are shown in Figs.\ \ref{inter}a, \ref{inter}b and \ref{inter}c 
 by solid lines. 
The ''true`` masses and scattering rates
\begin{equation}
\gamma_{opt}(\omega,T)-i\omega m^*_{opt}(\omega,T)/m_b
=\frac{\omega_{pl}^2}
{4\pi \sigma(\omega)}.
\label{epstrue}
\end{equation}
 and  the {\it chi-by-eye  \cite{numrec} } fit by the formula (\ref{padx}) are plotted 
in Figs.\ \ref{inter}a, \ref{inter}b and \ref{inter}c 
by dotted and  dashed lines respectively.

Note, that due to the frequency averaging, the quasipartical optical conductivity
(\ref{signr},\ref{sigsup}) is a slowly varying function in the mid-infrared 
region. In contrast,  the interband permeability near  transition edge 
(here, near the center of the Lorentzian) can changes rapidly. The second 
important feature of $\epsilon_{inter}(\omega)$ is its very weak temperature
dependence.

The mixture of the standard Drude term ($\gamma$=1000 cm$^{-1}$) with the  Lorentzian
(\ref{epsinter}) is shown in panel \ref{inter}a. This simple artefact
can be easily recovered, since it is temperature independent and the optical mass
changes too quickly.

The second and third  examples  are the half-by-half blends of 
the ''model 2`` ISB conductivities (T=100 K, panel \ref{inter}b;T=300 K, panel \ref{inter}c)
with the same $\epsilon_{inter}(\omega)$ Eq.\ (\ref{epsinter}).  It is more
complicated case, since the nonmonotonic behaviour could be screened by forthcoming
transitions. In this case  detailed quantitative analysis of both  the 
frequency and the temperature dependenc{\it ies} is needed. One has
 to pay the special attention to
the temperature dependencies of $m^*_{opt}(\omega\rightarrow 0,T)/m_b$, 
$\omega_{(m^*-1)/2}$ and
$\gamma_{opt}(\omega\approx 2\pi\lambda T,T)$.

 If the interband transitions
are wide spectral bands without  unqualified features,
their contribution to the permeability  is usually included in
 $\epsilon_{\infty}$. The effects caused by the underestimation of  its  
real part was considered by us in \cite{shulga91}.  Here we study
the problems produced by the use of Eq.\ \ref{idiotsigma} and by the 
underestimation of the imaginary part of $\epsilon_{\infty}$. Since the 
last quantity can be approximated by the simple constant only in the 
limited frequency range, for the sake of the analytical correct 
the model function was chosen in the form
\begin{eqnarray}
Im(\delta\epsilon_{\infty}(\omega)) & = & \left[\tanh{\left( \frac{\omega-2000}{5000}\right)}+
\tanh{\left( \frac{50000-\omega}{30000}\right)}\right]  \mbox{\ \ \ \ at $\omega>0$}\nonumber , \\
Im(\delta\epsilon_{\infty}(-\omega)) & = &  - Im(\delta\epsilon_{\infty}(\omega)) \nonumber , \\
Im(\delta\epsilon_{\infty}(0)) & = &  0 , \nonumber \\
Re(\delta\epsilon_{\infty}(\omega)) & = & 2.5 -\frac{2\omega}{\pi}\int_{0}^{\infty}
\frac{dxIm(\delta\epsilon_{\infty}(x))}{x^2-\omega^2}  .
\end{eqnarray}
In the mid infrared $(\delta\epsilon_{\infty}(\omega))\approx 4+2i$.

The model  frequency dependencies of the optical masses
and scattering rates 
\begin{equation}
\gamma_{opt}(\omega,T)-i\omega m^*_{opt}(\omega,T)/m_b
=\frac{\omega_{pl}^2}
{4\pi \sigma(\omega)-i\omega \tilde{\delta}(\epsilon_{\infty})}.
\label{epsinf}
\end{equation}
are shown in panel \ref{inter}d for
 $\tilde{\delta}(\epsilon_{\infty})=\delta(\epsilon_{\infty}(\omega))-1$ (solid lines),
 $\tilde{\delta}(\epsilon_{\infty})=\delta(\epsilon_{\infty}(\omega))-4$ 
(dashed-dotted lines),
 and  
$\tilde{\delta}(\epsilon_{\infty})=2i$ (dashed lines). The unperturbed 
$\gamma_{opt}(\omega,T)$ and $m^*_{opt}(\omega,T)/m_b$
(''model 2`` spectral function, $\lambda$=2, T=300 K) are plotted by 
dotted lines for comparison. As usually, the masses do not lie, and manifest
that  the ''unusual`` frequency dependence of the scattering rate is no more
than the artefact of the chosen interpretation procedure.

The early band structure calculations \cite{maksimov89} predicted, that the interband contribution
into the in-plane HTSC permeability is a real constant  between 10 and 15
 at $\omega\rightarrow 0$,
and becomes complex for $\omega>0.4 eV$. In the vicinity of the plasma edge
its   value is  4+2i. This information was used
by us for the successful description of the {\it normal state} optical properties 
of HTSC materials both in FIR and MIR spectral regions \cite{shulga91,jos1}.

In conclusion, the overlooked interband transitions produces the 
''unusual`` frequency dependence of the optical relaxation, which can
be nevertheless analysed properly  within the ISB model.

\subsection{Why is the first derivative from the experimental curve
better than the second one?}

Let us rewrite the valid at T=0 
Eqs.(\ref{goptT0},\ref{gamefT0}) in the differential form
\begin{eqnarray}
\alpha^2F(\omega)& = &\frac{1}{2\pi}\frac{d\gamma_{eff}(\omega)}{d\omega} 
\label{1prg} \\
\alpha^2F(\omega) & = & \frac{1}{2\pi}\frac{d^2(\omega\gamma_{opt}(\omega))}{d\omega^2}.
\label{2prg}
\end{eqnarray}
At a glance both Eqs.\ (\ref{1prg}) and (\ref{2prg}) look nice and simple. 
 But if one attempts  to apply them to an experimental data, the distinction
becomes evident. The point is that the experimental data are known only approximately.
A finite value of the signal to noise ratio (S/N) is the reason why the 
Eq.\ (\ref{2prg}) can not be applied to the FIR data. Typical value of the 
reflectivity S/N 
do not exceed 50, the multi-reflection techniques \cite{farn76} allow to reach S/N=2000.
 Let us adopt for simplicity, that $f(\omega)=\omega\gamma_{opt}(\omega)$ 
is sampled in equidistant points
$\omega_{i}$, $i=1,M$ and all values
$f(\omega_i)$ have the same errors 
\begin{equation}
\frac{f(\omega_i)}{\delta f(\omega_i)}=\left(\frac{S}{N}\right)_f.
\end{equation}

Note, that in the normal state at $T=0$ (when the application of 
(\ref{1prg}) and (\ref{2prg}) is correct) both 
 $f(\omega)=\omega\gamma_{opt}(\omega)$ and its first derivative 
$f^{\prime}(\omega)$ are monotonically
increasing functions, if a metal under consideration
is the ISB one.

The difference $f(\omega_{i+1})-f(\omega_i)$ has the order of 
$f(\omega_i)/M$. 
Its error is the   $\sqrt{\delta f^2(\omega_{i+1})+\delta f^2(\omega_i)}\approx 
\sqrt{2}\delta f(\omega_i)$. As a result the first derivative $f^{\prime}(\omega)$ has
the signal to noise ratio 
\begin{equation} 
\left(\frac{S}{N}\right)_{f^{{ \prime}}}=\frac
{f^{\prime}(\omega_i)}
{\delta f^{\prime}(\omega_i)}= \frac{f(\omega_{i+1})-f(\omega_i)}
{\sqrt{\delta f^2(\omega_{i+1})+\delta f^2(\omega_i)}}\approx
\frac{1}{\sqrt{2}M}
\left(\frac{S}{N}\right)_f,
\end{equation}
 reduced by the factor $K_1=\sqrt{2}M$ in comparison with
the original (S/N) in $f(\omega)$.
 Similar  the second derivative $f^{\prime \prime}(\omega)$ has
the signal to noise ratio 
\begin{equation} 
\left(\frac{S}{N}\right)_{f^{{ \prime\prime}}}=\frac
{f^{\prime\prime}(\omega_i)}
{\delta f^{\prime\prime}(\omega_i)}
= \frac{(f^{\prime}(\omega_{i+1})-f^{\prime}(\omega_i))^{\prime}}
{\sqrt{\delta f^{\prime 2}(\omega_{i+1})+\delta f^{\prime 2}(\omega_i)}}
\approx \frac{1}{2M^2}
\left(\frac{S}{N}\right)_f,
\end{equation} 
reduced by the factor $K_2=2M^2$. 

In  subsection \ref{favor} the four parameters formula (\ref{padx}) was promoted
as the universal fitting tool. $M=4$ means $K_2=32$. In another words, 
the normal  state  $\gamma_{opt}(\omega)$ has to have the original 
signal to noise ratio (S/N)=100, in order  to define {\it four} points   
in its second derivative   $\gamma_{opt}^{\prime\prime}(\omega)$ with accuracy of 30\%. 

As for K$_1$, this quantity grows reasonably.
 Of course, one can not measure the effective scattering rate $\gamma_{eff}(\omega)$
and use the formula (\ref{1prg}), nevertheless
 the title question is worth-while. The point is that the first derivatives from 
the superconducting state optical scattering rate  $\gamma_{opt}^{\prime s}(\omega)$
and the reflectivity $R^{\prime}(\omega)$ reproduce the input
spectral function $\alpha^2F(\omega)$\cite{allen71,farn76,joyce70,farn74}.
The  well structured  spectral function of lead 
obtained from the analysis of the first derivative of $A_s(\omega)-A_n(\omega)$
looks convincing
(at least for me) due to the reported value of S/N=6000.

 Analysing Eq.\ (\ref{2prg}) we arrive at  the same conclusions  as discussed above.
The normal state optical conductivity provides us with restricted   
 information  similarly  as other physical methods.  The only
way to extract the all required material parameters values 
 is the analysis of different measurements
 in frame of the {\it same}  model.

\section{Superconducting state optical conductivity}
\label{super}
\subsection{The inversion procedure}

\subsubsection{Adaptivity, what does it mean?}
\label{awdm}
Let us consider from a general  point of view the problems, which one 
has to take in mind, when one try to reconstruct 
 a spectral function $\alpha^2F(\omega)$
from  experimental reflectivity data. This analysis should be performed
 in frame of a  model which is based on an appropriate formalism. 
Thus, it means, that the results obtained are model dependent.
The ideal, but rare situation  occur, when  the
exact analytical solution of the inverse problem is available 
\begin{equation}
\alpha^2F(\omega)=f(R(\omega),\omega)
\label{onefor}
\end{equation}
where $f$ is a known function, which could depend on $\alpha^2F(\omega)$.
In the last case the numerical solution can be found by iterations.
 
But what should we do, if such a function $f$ is not available by reason of 
the complexity of the direct formalism. In this case we shell  use 
the numerical {\it adaptive strategy}, based on the substitution of equations of the 
direct formalism by  the {\it adaptive} formula(s), suitable 
for the numerical solution of ill-posed problems.   
 Since it is a strategy, let us consider how does the adaptivity   work.

Generally speaking, one can use {\it any} adaptive formula. For example,
one can consider the $\alpha^2F(\omega)$ as an input function and the
reflectivity as an  output function for some formalism. Then one could write
\begin{equation}
R(\omega)=\int d z K(\omega-z)\alpha^2F(z),
\label{ad0}
\end{equation}
or in the Fourier space 
\begin{equation}
{\cal F}(R(x))={\cal F}(K(x)){\cal F}(\alpha^2F(x)),
\label{furad}
\end{equation}
where ${\cal F}(f(x))$ is the Fourier image 
of the function $f$ and  K(x) is an for the time being unknown kernel.
 K(x) could be determined as follows.
\begin{itemize}
\item
At first, one has to substitute 
a {\it trial} spectral function $\alpha^2_{trial}F(\omega)$ into the equations
of the main formalism (Eliashberg equations + Nam-Rainer formula)
 and obtains a {\it trial} reflectivity 
$R_{trial}(\omega)$. 
\item
Next, one should substitute $\alpha^2_{trial}F(\omega)$ and
the {\it calculated} $R_{trial}(\omega)$ into the adaptive formula (\ref{furad})
and solve Eq.\ (\ref{furad}) for K$_1(\omega)$.
  The subscript 1 means ''the first iteration``.
\item
Finally, one has to substitute the {\it now known} kernel K$_1(\omega)$ and the 
{\it experimental} R$_{exp}(\omega)$ into the same adaptive formula (\ref{furad}),
and solve it for  $\alpha^2_{1}F(\omega)$.
\end{itemize}
 So, on each iteration
we use the model formalism ones and the adaptive formula twice. 
At first, one have to calculate $R_{k-1}(\omega)$ using $\alpha^2_{k-1}F(\omega)$,
as input function.
At second, one should find  the 
adaptive function(s)
\begin{equation}
{\cal F}(K_{k}(x))=\frac{{\cal F}(R_{k-1}(x))}{{\cal F}(\alpha^2_{k-1}F(x))},
\label{ad1}
\end{equation}
and finally  calculate the spectral function
\begin{equation}
{\cal F}(\alpha^2_{k}F(x))=\frac{{\cal F}(R_{exp}(x))}{{\cal F}(K_{k}(x))}.
\label{ad2}
\end{equation}
The presented ''trick`` is quite general.
One could replace analogously, for example, $R(\omega)$ in (\ref{ad0}-\ref{ad2}) 
by  the tunnelling density of states
  $N(\omega)$ or the ARPES spectral function $A(\omega)$, and obtain
 adaptive methods for the 
tunnelling and electronic spectroscopies respectively,
 which is expected can work, with any formalism. 
 
As mentioned above, the experimental data are known only approximately,
therefore the inverse problem is ill-posed. 
Up to now the solution of 
ill-conditioned problems is the art more than the routine procedure.
Since  one can use any adaptive formula(s), it will be pragmatic for him/her/AI
to choose it from the short (alas!) list of well investigated
by numerical mathematics equations (see, for example,  sec. 18.4-18.7 in \cite{numrec}
 and references within).

On this understanding 
the discussed above  convolution adaptive formula
is one of the best possible selections, 
since the Fourier transformation is linear and unitary.

\subsubsection{The inversion method for the {\it s-}wave ISB reflectivity. }

The inverse problem for the superconducting state 
optical absorptivity of Pb was successfully solved by B.\ Farnworth and T.\ Timusk 
\cite{farn76} within the weak coupling ISB model, based on the theory
proposed by P.B. Allen \cite{allen71}.  The difference between 
weak coupling theoretical calculations and experimental data \cite{joyce70} 
was analysed in detail in \cite{allen71}. It was pointed, that:
\begin{itemize}
\item The first derivative from the superconducting state optical scattering rate
 contains the component, proportional to the input spectral function
$\alpha^2F(\omega)$ due to the influence of coherent factors (\ref{cohfac}).
\item  The weight $\omega/2\Delta_0$ of this component is strongly underestimated.
\item If one substitutes the solutions of  (\ref{ImEl}) into the dominators
of (\ref{sigsup}) and keeps BSC values $\tilde{\omega}=\omega$ and 
$\tilde{\Delta}=\Delta_0$ for the   coherent factors (\ref{cohfac}), one obtains 
the optical relaxation, close to the one, given by the weak coupling Allen theory.
\end{itemize}
Below we will see that ISB model based on the Nam formalism have qualitatively
the same properties.   
 
There are several reasons, why the following convolution adaptive 
formula 
\begin{eqnarray}
\frac{\gamma_{opt}^s(\omega)}{d\omega}
\theta(\omega) & = &2\pi\int d z K(\omega-z)\alpha^2F(z)\theta(z) 
\label{swaf} \\
\theta(x)& = & 1 \mbox{\ if\ } x\ge 0, \mbox{\ else\ } \theta(x)=0 \nonumber   
\end{eqnarray}
is suitable. At first, we segregate the  ill-posed part of the inverse
{\it s-}wave problem (taking of the first derivative) into the separate step.
At second, the  low and high frequency values of $\gamma_{opt}^{'s}(\omega)$
coincide (=0), it is important for the FFT algorithm. At third,
the obtained kernel $K(\omega)$ reminds us the instrumental resolution
function of a badly adjusted spectrometer, allowing to profit by  the 
applied spectroscopy experience.  If one thinks in terms of 
the transfer functions, Laplace transformation adaptive formula could be more suitable.

\begin{figure}
\centerline{\hbox{   
\hspace{-2.7cm} 
\psfig{figure=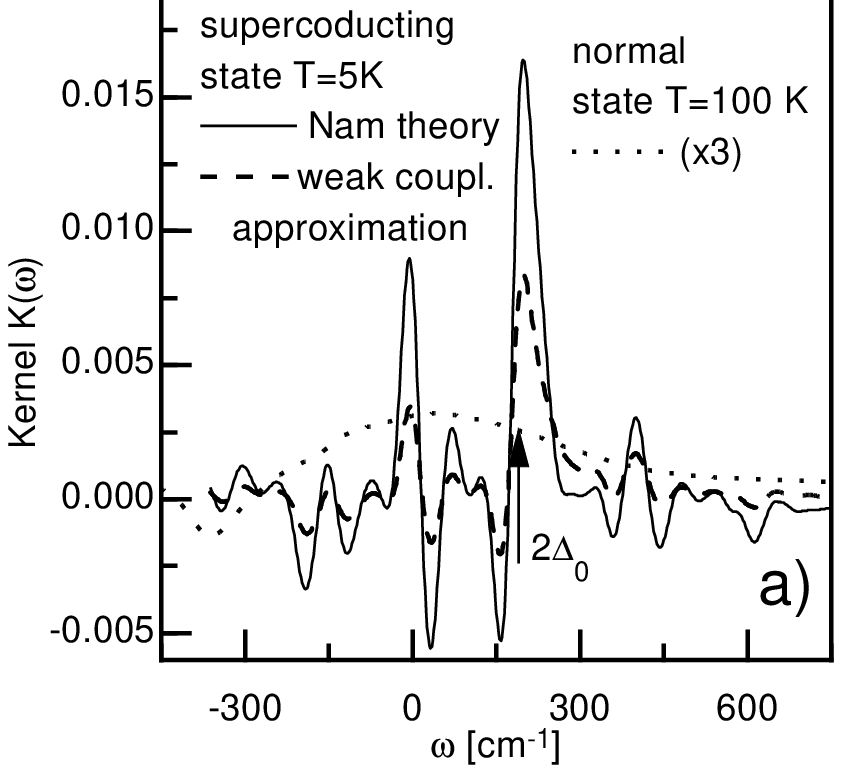,width=6.0cm,height=6.0cm}
\hspace{0.8cm} 
\psfig{figure=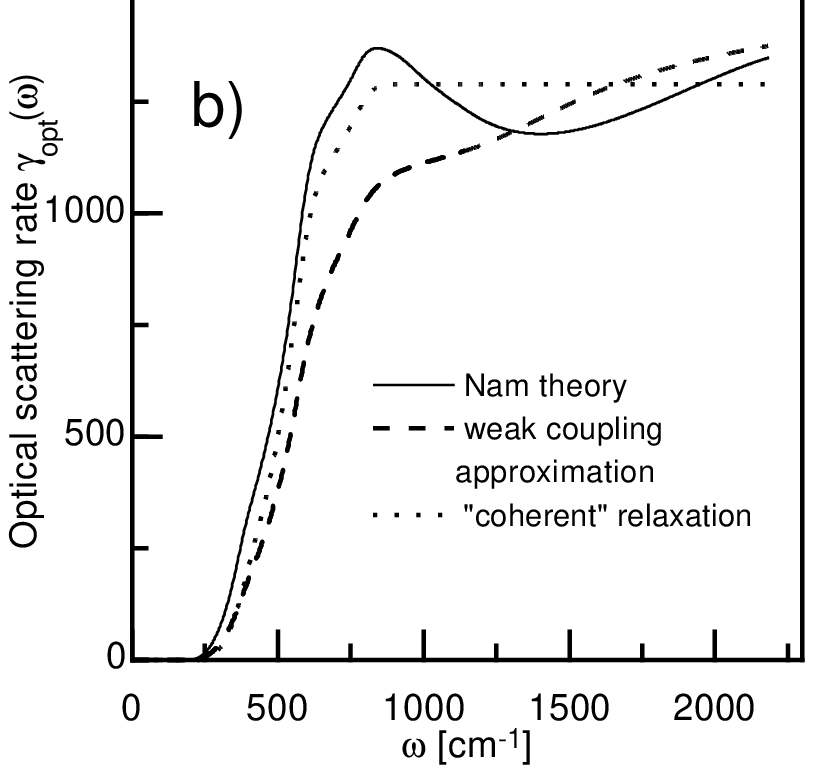,width=6.0cm,height=6.0cm}
}}
\vspace{1cm}
\caption{Panel (a) shows the frequency dependence of the 
superconducting state kernels  $K(\omega)$ (solid and dashed lines) in comparison
with the normal state one (dotted line). The source  superconducting state 
optical scattering rates, 
calculated  in the strong (solid line),
and weak (dashed line) coupling approximations are shown in panel (b).
The convolution of the input spectral function with the master peak
 (=''coherent`` relaxation) is drown in panel (b) by the dotted line.}
 \label{svertka}
 \end{figure}

The equation (\ref{swaf}) was solved  with respect to  
$K(\omega)$ using optical scattering rates {\it calculated}
 in strong (solid line, Fig.\ \ref{svertka}b) 
and weak (dashed line, Fig.\ \ref{svertka}b) coupling approximations and input
''model 2`` $\alpha^2F(\omega)$. The results
are plotted by the same type lines in Fig.\ \ref{svertka}a. No doubt, all features of 
the kernel are important for the recovery of the precise peak amplitudes
and energy positions. Nevertheless, the general shape and main physics of 
$\gamma_{opt}^{s}(\omega)$ are 
defined by the {\it master} peak, located just above $\omega=2\Delta_0$.
The arrow, pointing in Fig.\ \ref{svertka}a
the frequency position of $2\Delta_0$, is located within this peak. The master peak
income 
is illustrated by dotted line in Fig.\ \ref{svertka}b. It was calculated 
as follows. 
 
\begin{itemize}
\item All collateral maxima and minima of the shown by solid line  $K(\omega)$
 were rejected and artificial 
kernel $K_{coh}(\omega)$ was set to zero anywhere besides the 
master peak frequency region, where it coincides with  $K(\omega)$.
\item $K_{coh}(\omega)$ was convoluted with the input spectral function
$\alpha^2F(\omega)$ and integrated over $\omega$
\item Since the discussed phenomenon is governed by the coherence factors,
the resulting ''scattering rate`` (dotted line, Fig.\ \ref{svertka}b) is labelled
as {\it coherent relaxation}.
\item The simple frequency dependence of the  coherent relaxation,
reminding us the properties of the quasiparticle effective scattering
rate $\gamma_{eff}(\omega)$, allow to enunciate the 
Visual Accessibility procedure, described in forthcoming section.
\end{itemize}
The normal state kernel (multiplied on 3) is shown by the dotted line on 
Fig.\ \ref{svertka}a for comparison. The pessimistic conclusion, made in the previous
chapter, can be couched in current terms:
 the (induced by the frequency  and temperature averaging)
width of $K(\omega)$ is too large to resolve the fine structure of the spectral function.

The domination of the master peak makes the choice of the trial $\alpha^2F(\omega)$
on the first iteration easy. It can be the normalised $\gamma^{\prime,s}_{opt}(\omega)$
 or $R^{\prime}_{s}(\omega)$. 
Note, that the master peak is asymmetric. The final frequency resolution is defined by 
the small value of its left half-width.
 
The self-consistent determination of the value of the superconducting gap
is theoretically possible in frame of adaptive approach, but strongly not recommended.
Following this way one transforms the weakly nonlinear task to the strongly nonlinear
problem. For the case of ill-conditioned equations it is the wrong way.
The value of $\Delta_0$ has to be known {\it a priori}. The related issue is the
difference between $\alpha^2F(\omega)$ and $\alpha^2_{tr}F(\omega)$. 
Substituted to the Eliashberg equations transport spectral function 
$\alpha^2_{tr}F(\omega)$ lands us in difficulty, since the obtained value of the 
 gap can be lower than the known $\Delta_0$.  In my opinion, for the sake of simplicity
 in this case 
one might sacrifice the unknown (from the first principle calculations)
Coulomb pseudopotential and use $\mu^*$
 as a simple fitting parameter.

In conclusion, the {\it adaptive} 
 method of 
 the {\it exact  numerical inversion} of the {s-wave} superconducting state 
 Nam equations is presented. In my opinion, one day the FIR spectroscopy will provide
not an alternative to the tunnelling, but a powerful independent  tool
 for the determination
of $\alpha^2F(\omega)$ in {\it s-}wave superconductors.

\subsection{Rules of Visual Accessibility procedure}

The frequency dependence of the  superconducting state
ISB reflectivity $R_s(\omega)$ in the local limit has
 the distinctive forms. 
This features are determined by the physical properties of the 
optical scattering rate $\gamma_{opt}^s(\omega)$. In this section we consider
our simplified approach (Visual Accessibility)  which allow us 
 to  analyse ''by eye`` the experimental reflectivity curve and get  qualitatively 
the same results as the  sophisticated method, described in the previous section,
able to obtain quantitatively.
 
\subsubsection{Optical scattering rate and reflectivity}

As far as  the physics is determined by the $\gamma_{opt}(\omega)$, but
the measured  quantity is $R(\omega)$,
 let us derive a simple relation between  
them.
In view of a high value
of the  permeability $\epsilon(\omega)\gg 1$ one can
expand the formula 
\begin{equation}
R(\omega)=\left |\frac{1-\sqrt{\epsilon(\omega)}}{1+\sqrt{\epsilon(\omega)}}\right |^2
\end{equation}
in powers  of $1/\sqrt{\epsilon(\omega)}$. 
In view of Eq.\ (\ref{eqw}), substituting $1+\lambda$ instead of  
$m^{*}_{opt}/m_b$, one arrives
the following approximate relation 
 \begin{equation}
R(\omega)\approx 1-\frac{2\gamma_{opt}(\omega)}{\omega_{pl}\sqrt{1+\lambda}}
\end{equation}
valid at $\omega\le\Omega_{max}+2\Delta$ both in the normal
($\Delta$=0) and superconducting ($\Delta$=$\Delta_0$) states,
except for small frequencies $\omega < \gamma_{opt}(0)/(1+\lambda)$,
where Hagen-Rubens approximation is suitable.

\begin{figure}
\centerline{\hbox{   
\hspace{-2.3cm} 
\psfig{figure=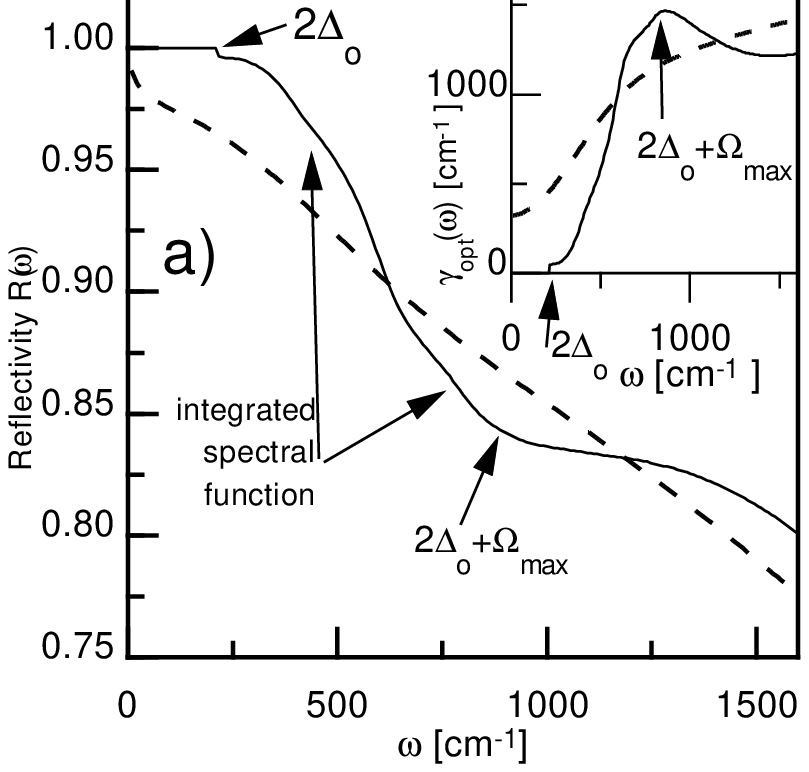,width=6.0cm,height=6.0cm}
\hspace{0.8cm} 
\psfig{figure=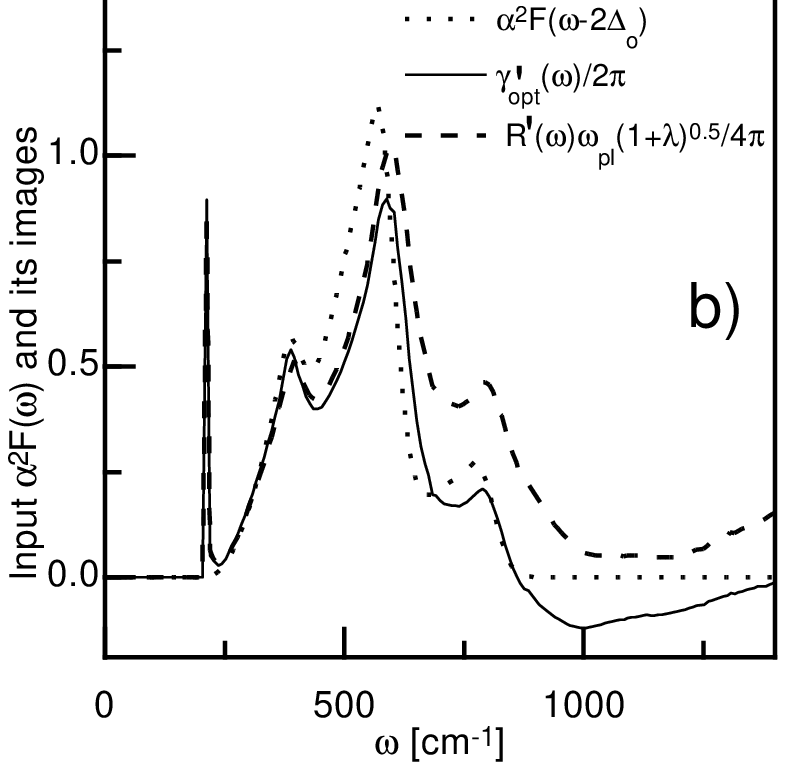,width=6.0cm,height=6.0cm}
}}
\vspace{1cm}
\caption{Panel
 (a) shows the frequency dependence of the superconducting state 
reflectivity
 $R_s(\omega)$
(T=10 K, solid line) in comparison with
the  normal state 
 reflectivity $R_n(\omega)$ (T=100K, dashed line).
Both curves were   calculated using ''model 2`` spectral 
function.
The  frequency dependencies of the optical scattering rates 
$\gamma_{opt}^s(\omega)$ and  $\gamma_{opt}^n(\omega)$ 
are shown in the inset. Panel (b) shows the frequency dependencies 
of the  first derivatives from 
the  optical scattering rate $\gamma^{\prime s}_{opt}(\omega)/2\pi$ 
and  from the normalised reflectivity 
$R^{\prime}(\omega)\omega_{pl}(1+\lambda_0)^{0.5}/4\pi$
in comparison with the shifted on $2\Delta_0$ input spectral function
$\alpha^2F(\omega-2\Delta)$.  One can see, that due to domination of the 
coherent 
relaxation in the frequency region between $2\Delta_0$ and  $2\Delta_0+\Omega_{max}$,
$R^{\prime}(\omega)\omega_{pl}(1+\lambda_0)^{0.5}/4\pi$ and 
$\gamma^{\prime s}_{opt}(\omega)/2\pi$  reproduces well the peaks of the 
spectral function. The important features are indicated in panel (a). One 
can see, that the absorption onset at $2\Delta_0$, the frequency dependence
of the coherent relaxation between  $2\Delta_0$ and  $2\Delta_0+\Omega_{max}$,
and its upper energy bound  at   $2\Delta_0+\Omega_{max}$ are visually accessible.
The term ''integrated spectral function`` means, that the first derivative over $\omega$
from this part of the spectrum reproduces $\alpha^2F(\omega-2\Delta_0)$. }
\label{facefig}
 \end{figure}

Fig.\ \ref{facefig}a shows the frequency dependence of the superconducting state 
reflectivity
 $R_s(\omega)$
(T=10 K, solid line) in comparison with 
the  normal state 
 reflectivity $R_n(\omega)$ (T=100K, dashed line). 
Both curves have been  calculated using the ''model 2'' spectral function.
 The  frequency dependencies of the optical scattering rates 
$\gamma_{opt}^s(\omega)$ and  $\gamma_{opt}^n(\omega)$ 
are shown in the inset.  One observes that $R(\omega)$ looks like the mirror image of 
 the appropriate $\gamma_{opt}(\omega)$, reproducing all important features.

\subsubsection{The visual accessibility criteria for {\it s-}wave superconductors}
\label{vac}
Let us summarise  our knowledge about the {\it s-}wave superconducting state optical 
relaxation rate and the local (London) limit reflectivity (see Fig.\ \ref{facefig}). 
The following features are visually accessible. 
\begin{itemize}
\item 
The doubled gap value 2$\Delta_0$ coincides with the absorption edge.
If the $\omega<2\Delta_0$, the reflectivity R($\omega$)=1 and 
$\gamma_{opt}^s(\omega)$=0.
\item The impurity scattering gives a sharp decrease of the reflectivity
just above  2$\Delta_0$. The magnitude of the "jump" is proportional to
 $\gamma_{imp}$.  $\gamma_{opt}^s(\omega)$ jumps approaching  $\gamma_{imp}$
from above. 
\item
 In superconducting state 
the coherent   relaxation dominates.  Hence, the first derivatives from 
  $\gamma^{\prime s}_{opt}(\omega)$  and  from  R($\omega$)
reproduce the shape of the spectral function (see Fig. \ref{facefig}b). 
\item 
Since the frequency dependence of the  coherent relaxation rate
grows  faster, than the frequency averaged normal state one,
 and is fulfilled its maximum value at 
 the frequency $\omega=2\Delta_0+\Omega_{max}$, at this point the
ratio of the $\gamma^{ s}_{opt}(\omega)/\gamma^{n}_{opt}(\omega)$  has a maximum,
while   the reflectivity ratio $R_{ s}(\omega)/R_{n}(\omega$) exhibits a minimum.
\end{itemize}

\subsection{Visual accessibility and separable  $\mbox{\it d-}$wave model}

In this section we consider briefly, how the procedure described in section 
\ref{vac} should be corrected in order to make the main features of the {\it d-}wave
ISB reflectivity visually accessible.

\subsubsection{Separable  {\it d-}wave model}

\begin{figure}
\centerline{\hbox{   
\hspace{-2.3cm} 
\psfig{figure=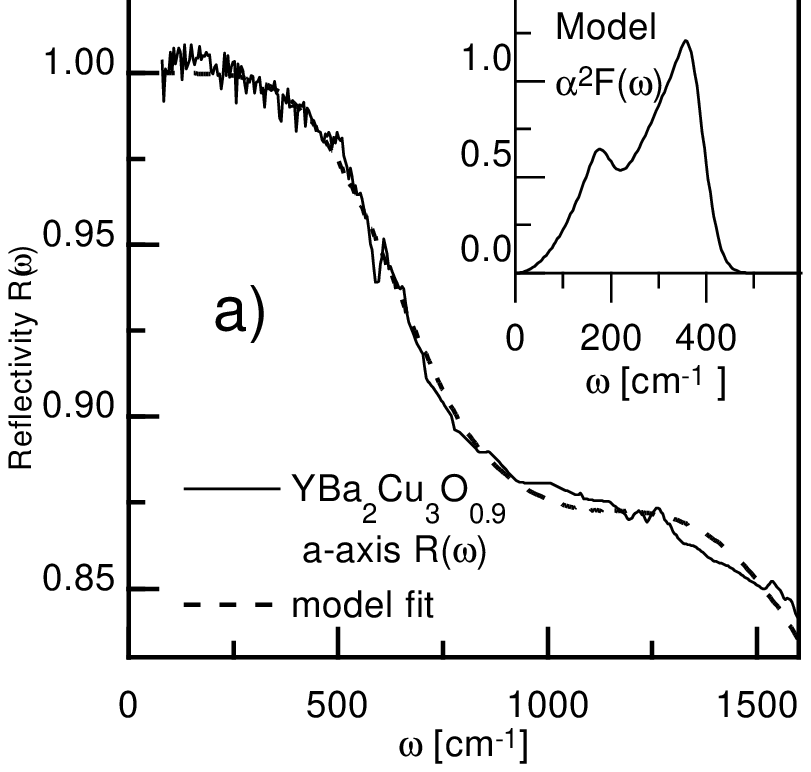,width=6.0cm,height=6.0cm}
\hspace{0.8cm} 
\psfig{figure=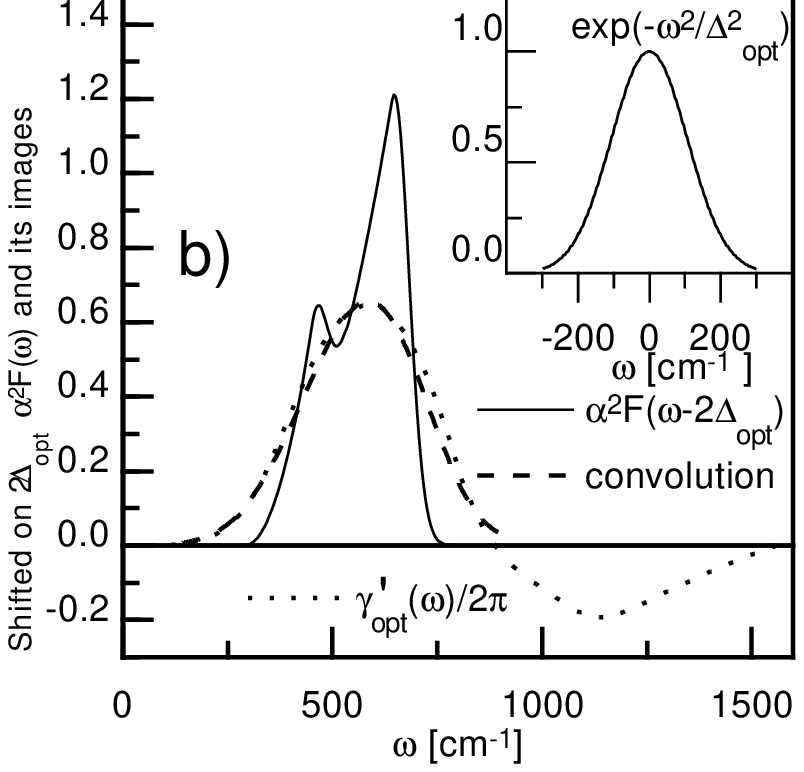,width=6.0cm,height=6.0cm}
}}
\centerline{\hbox{   
\hspace{-2.3cm} 
\psfig{figure=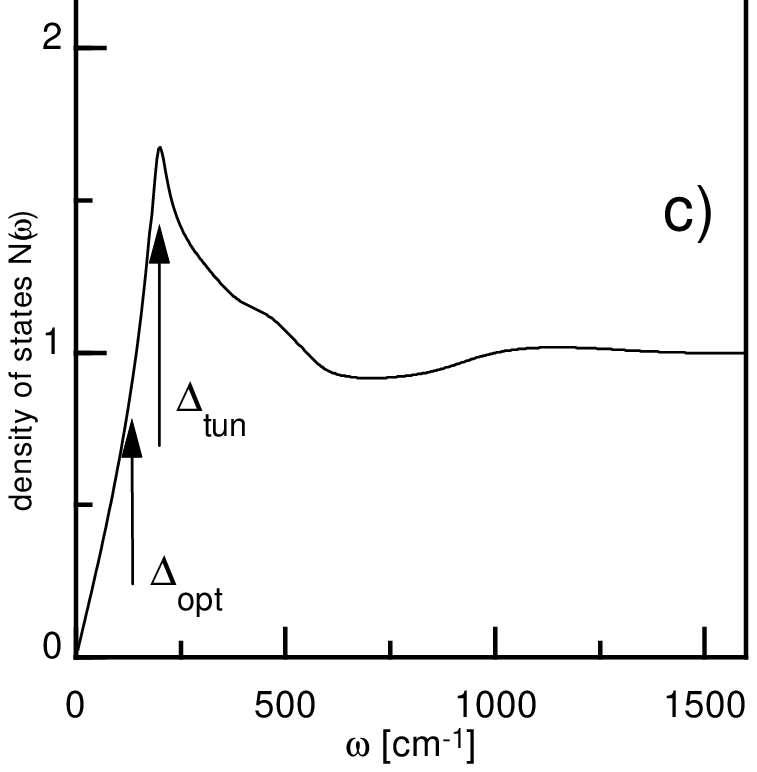,width=6.0cm,height=6.0cm}
\hspace{0.8cm}     
\psfig{figure=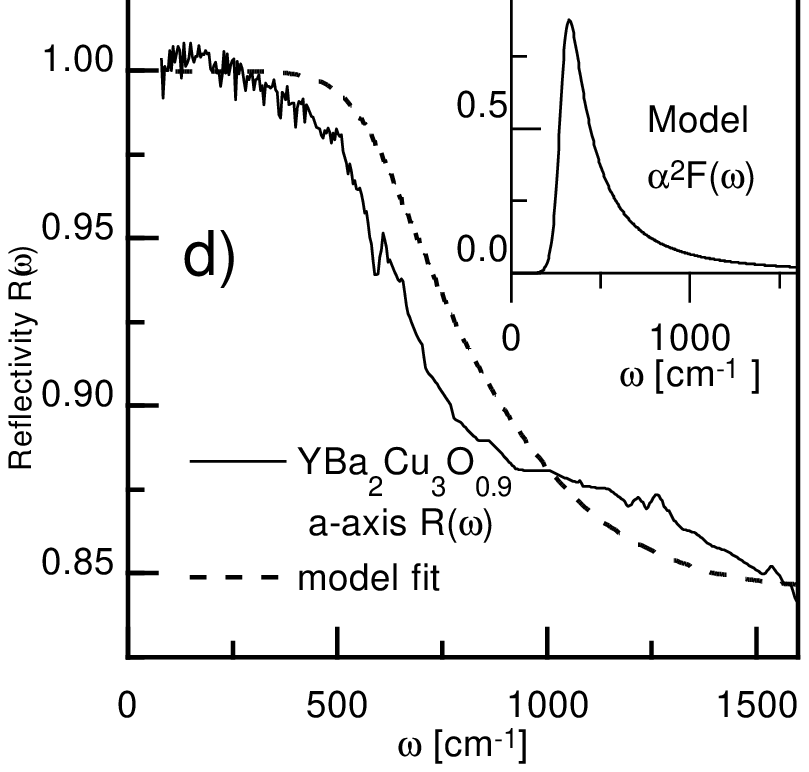,width=6.0cm,height=6.0cm}
}}
\vspace{1cm}
\caption{Panels (a)  shows the frequency dependency
 of the  reflectivity R$_s(\omega)$ (dashed lines)
 calculated 
within the separable {\it d-}wave
model in comparison with the experimental data (solid line, Schutzmann {\it et al}
 {\it Phys. Rev.} {\bf B46} 512 (1992)). 
The  spectral functions
$\alpha_0F(\omega)$ are shown in the  insets. The  coupling constants were chosen 
to obtain the value of the  tunnelling gap $\Delta_{tun}$=200 cm$^{-1}$. 
The values of the plasma frequencies $\omega_{pl}$
and the low-frequency values of the non-metallic  part of dielectric
permeability $\epsilon_{\infty}$ were obtained by the least-square fit.
Panel (b) shows the frequency dependence of the first derivative from the calculated
 optical scattering rate
$\gamma^{\mbox{\ }\prime}_{opt}(\omega)$ (dotted line) in comparison with the
 shifted on 2$\Delta_{opt}$  input
spectral function $\alpha_0^2F(\omega-2\Delta_{opt})$(solid line). The convolution
of the $\alpha_0^2F(\omega-2\Delta_{opt})$ with the Gauss distribution function
(shown in inset (b)) is plotted by the dashed line.
Panel (c) shows the frequency dependence of the calculated tunnelling density of states
N($\omega$), the frequency positions of the tunnelling gap $\Delta_{tun}$ and the
optical gap  $\Delta_{opt}$. Panel (d) shows the frequency dependencies of the 
 reflectivity R$_s(\omega)$ (dashed lines), calculated using 
caudate spectral function (inset),
in comparison with the experimental data.} 
 \label{fig3}
 \end{figure}
\noindent

Formally speaking, the frequency dependencies of the isotropic and the
anisotropic parts of the spectral function are not obliged to coincide.
At the same time,  if the shapes differ substantially, the density of states
(\ref{nscm1}) looks strange. So, for the sake of tunnelling, the model
spectral function was chosen in the simple form  
\begin{equation}
\alpha^2F(\omega,\varphi)=\alpha^2_0F(\omega)(1+2gcos(2\varphi)cos(\varphi^{\prime}))
\label{alfd}
\end{equation}
If one substitutes the function (\ref{alfd})
into the $s+id$ ISB Eliashberg equations, one finds that at $g>1$ the {\it d-}wave
component entirely dominates. Hence, instead of the 
sophisticated $s+id$ approach,
 we can use for our analysis the pure {\it d-}wave
model  described in detail in \cite{schachiger97}. 
For the sake of simplicity 
we restrict ourselves here to the  consideration 
of the clean limit $\gamma_{imp}=0$ and
set the  Coulomb pseudopotential $\mu^*(\varphi,\varphi^{\prime})$ to zero.  
The value of the parameter g was chosen little a bit larger than unity, since
the normal state properties (defined in frame of {\it this} model by 
$\alpha^2_0F(\omega)$) manifest  the strong renormalization due to
the phonon-electron interaction.

It is interesting, 
that it is possible after minor rehash  to use the available 
{\it s-}wave computational programs 
for the
strong coupling {\it d-}wave calculations. 
The clean limit version of the separable {\it d-}wave  model keeps the convolution
structure of the set (\ref{ImEl}). One has to substitute 
the angle averaged density of states
\begin{equation}
\left<\frac{\tilde{\omega}(y)}{\sqrt{\tilde{\omega}(y)^2-\tilde{\Delta}^2(y)}}
\right>_{\varphi}=
\frac{1}{\pi}\int_0^{\pi} 
\frac{\tilde{\omega}(y)d\varphi}
{\sqrt{\tilde{\omega}^2(y)-2\tilde{\Delta}^2(y)cos^2(2\varphi)}}
\label{nscm1}
\end{equation}
instead of $N(\omega)$ (\ref{Notom}) in  (\ref{ImEl}), and
the angle averaged density of pairs
\begin{equation}
\left<\frac{\tilde{\Delta}(y)}{\sqrt{\tilde{\omega}(y)^2-\tilde{\Delta}^2(y)}}
\right>_{\varphi}=
\frac{1}{\pi}\int_0^{\pi} 
\frac{2\tilde{\Delta}(y)cos^2(2\varphi)d\varphi}
{\sqrt{\tilde{\omega}^2(y)-2\tilde{\Delta}^2(y)cos^2(2\varphi)}}
\label{nscm2}
\end{equation}
instead of $D(\omega)$ (\ref{Dotom}) in  (\ref{ImEl}). Similarly, the conductivity
can be calculated using the Nam formula (\ref{sigsup}). For this purpose 
one should  substitute  
$\Delta_d(x)=\sqrt{2}\Delta(x)\cos{(2\varphi)}$  instead of the 
$\Delta_s(\omega)$ into the element of integration in  (\ref{sigsup}) 
and averages the obtained $\sigma(\omega,\varphi)$ over the angle $\varphi$. 

Fig.\ \ref{fig3}a shows the  
model  reflectivity R$_s(\omega)$, calculated in the frame of the separable
 {\it d-}wave model (dashed line)
in comparison with the experimental data \cite{renk92} (solid line). 
The  spectral function
$\alpha_0F(\omega)$ is shown in the  inset in Fig.\ \ref{fig3}(a),
 its coupling constant $\lambda$
 was chosen 
to reproduce  the value of the {\it tunnelling} gap $\Delta_{tun}$=200 cm$^{-1}$ 
derived from  the tunnelling  spectra \cite{blin}. 
Note, that it was the same ''model 2`` \cite{shulga91} curve, used above,
 but without any high energy peak. The value of the plasma frequency $\omega_{pl}$
and low frequency value of the non-metallic  part of dielectric
permeability $\epsilon_{\infty}$ were obtained by the least-square fit.

Fig.\ \ref{fig3}b shows the first derivative from the {\it d-}wave optical scattering rate
$\gamma^{\mbox{\ }\prime}_{opt}(\omega)$ (dotted line) in comparison with the
 shifted on 2$\Delta_{opt}$  input
spectral function $\alpha_0^2F(\omega-2\Delta_{opt})$(solid line).
 Here we define the 
{\it optical} gap  $\Delta_{opt}\approx\Delta_{tun}/\sqrt{2}$ as the physical
quantity, with defines the shift of the spectral function $\alpha_0^2F(\omega)$
in FIR and tunnelling spectroscopy. The distinction between  $\Delta_{opt}$
and $\Delta_{tun}$  arises due to the angle averaging ($<...>_\varphi$) 
both in the Eliashberg equations
and in the conductivity calculation.  The tunnelling gap $\Delta_{tun}$ (see 
Fig.\ \ref{fig3}c) 
is the maximum value of the {\it d-}wave gap
\begin{equation}
\Delta^d(\varphi)=\sqrt{2}\Delta cos(2\varphi)
\end{equation}
and the optical one  is its mean square angle average. 
Note, that the dip in the density of states 
N($\omega$) (see Fig.\ \ref{fig3}c) arises at $\Delta_{opt}+\Omega_{max}$ and
the electron-boson region starts in $N(\omega)$ at $\Delta_{opt}<\Delta_{tun}$. 
  It means, that the 
low-energy excitations (if they contribute to the spectral function) will degrade 
the sharp peak in the tunnelling density of states. 

If one substitutes 
$\gamma^{\mbox{\ }\prime}_{opt}(\omega)$  in Eq.\ \ref{swaf}, one finds
that coherent relaxation still dominates, but the main peak is substantially 
wider and can be approximated by a Gaussian
function $Qexp(-\omega^2/\Delta^2_{opt})$ ((Fig.\ \ref{fig3}b, inset).
 The function
\begin{equation}
C(\omega)=\alpha^2_0F(\omega-2\Delta_{opt})\otimes\frac{1}{\sqrt{\pi}\Delta_{opt}}
\exp{\left(-\frac{\omega^2}{\Delta^2_{opt}}\right)}
\label{coto}
\end{equation}
is plotted by the dashed line in the Fig.\ \ref{fig3}b and agrees well with the 
$\gamma^{\mbox{\ }\prime}_{opt}(\omega)/2\pi$. Here $\otimes$ means 
a convolution.
Note, that the coincidence of the prefactor Q with 
the $1/\sqrt{\pi}\Delta_{opt}$
 is come-by-chance. 

In my opinion, the additional ill-posed step, the {\it real} deconvolution
of the Gaussian can not be performed using the data with 
 reasonable value of (S/N), since the
spectral function in these compounds have  half-width of the order of $\Delta_{opt}$.
On the other side, if one (by chance) will find  that  
$\gamma^{\mbox{\ }\prime}_{opt}(\omega)$ or $R^{\prime}(\omega)$ 
are structured, it means, that the ISB separable {\it d}-wave model is not suitable for
HTSC and have to be replaced.

Strictly speaking, the {\it d-}wave reflectivity is less than unity at any frequency.
At the same time,  there is an onset of the absorption
 at some $\omega_{\Delta}$,   
 where R($\omega$) becomes visually  different from unity.
 Since the convolution C($\omega$)
 is wider 
than the input spectral function $\alpha^2_0F(\omega)$, the position of the onset 
$\omega_{\Delta}<  2\Delta_{opt}$.  For the  model 
spectral function shown in the inset in  Fig.\ \ref{fig3}a
 $\omega_{\Delta}\approx  1.4\Delta_{opt}\approx \Delta_{tun}$. Generally speaking, 
the selected boson modes could  be responsible for the {\it d-}wave pairing. In this case
the position of the absorption onset will be between  $\Omega_{min}+\Delta_{opt}$
and  $\Omega_{min}+\Delta_{tun}$, where $\Omega_{min}$ is the low frequency bound
of the electron-boson interaction function. 
Similarly, the point where the coherent relaxation rate reaches its maximum,
will be shifted by  $\Delta_{opt}$ towards the high frequencies in comparison
with its position in the {\it s-}wave superconductors. As a result, the reflectivity
ratio  $R_{ s}(\omega)/R_{n}(\omega$) exhibits a minimum and the scattering rate
ratio
 $\gamma^{ s}_{opt}(\omega)/\gamma^{n}_{opt}(\omega)$ has a maximum
approximately at   $\Omega_{max}+2\Delta_{tun}$. 
 
\subsubsection{The visual accessibility criteria for {\it d-}wave superconductors}
\label{vacd}

In  ISB {\it s-}wave metals at the same frequency $\omega=\Delta_0$
the density of states exhibits three important features. At first, in this point
$N(\omega)$ has a maximum. At second, since the averaged over {\bf k} value
of $\Delta_0$ coincides in this case with itself, 
the density of states at $\omega=\Delta_0$
changes  most rapidly.  At third, for $\omega<\Delta_0$,
the value of $N(\omega)$ is equal to zero.  In {\it d-}wave superconductors
this structural feature has a different energy  position. One could denote 
\begin{itemize}
\item 
the position of the maximum in the density of states $\Delta_{tun}$ 
 as the {\it tunnelling} gap;
\item 
the value of the shift 2$\Delta_{opt}$  of the image  of the spectral function  
in the first derivative
from the optical relaxation rate $\gamma^{\mbox{\ }\prime}_{opt}(\omega)$ 
 as the {\it optical} gap , 
\item 
the frequency $\omega_\Delta$ 
 where R$_s$($\omega$) becomes visually  distinct from unity as the 
onset of the absorption
or as the {\it absorption} gap.  
\end{itemize}
The approximate relations between the gaps are 
$\Delta_{tun}\approx\omega_{\Delta}\approx\sqrt{2}\Delta_{opt}$.
Similar to the description of the visually accessible features
 made in subsection \ref{vac} for a {\it s-}wave metals,  we can
write  the same criteria for the separable {\it d-}wave model. Note, 
that in comparison
with the {\it s-}wave case, all presented below itemised statements 
are valid approximately only.
 
\begin{itemize}
\item 
The single  gap value $\Delta_{tun}$ coincides with the absorption edge.
If the $\omega<\Delta_{tun}$, the reflectivity R($\omega)\approx$1 and 
$\gamma_{opt}^s(\omega)\approx$0.
\item The impurity scattering does not give the sharp decreasing of the reflectivity
just above the $\Delta_{tun}$. Instead, it decreases the gap itself.
\item
 The coherent  electron-boson scattering dominates in the superconducting state.
The first derivatives from 
the  $\gamma^{\prime s}_{opt}(\omega)$  and  from the R($\omega$)
reproduce the spectral function shifted on 2$\Delta_{opt}$ and 
{\it convoluted} with the
Gauss distribution function $exp(-\omega^2/\Delta_{opt}^2$)\ (see Fig. \ref{fig3}b).
\item 
Since the frequency dependence of  the coherent relaxation rate
grows  faster than the normal state one,  and reaches its maximum value when
 the frequency $\omega\approx 2\Delta_{tun}+\Omega_{max}$, at this point the
ratio of the $\gamma^{ s}_{opt}(\omega)/\gamma^{n}_{opt}(\omega)$ will exhibit a maximum
and  reflectivities ratio $R_{ s}(\omega)/R_{n}(\omega$) will exhibit a minimum.
\end{itemize}

Finally,  let us make the simple acceptance test for the presented above 
visual accessibility  procedure.  The test spectral function has a long history.
At that old time, when people did   not distinguish  between the 
{\it effective} scattering rate $\gamma_{eff}(\omega)$ and  the {\it optical} one
 $\gamma_{opt}(\omega)$, Collins {\it et al}   \cite{collins87} interpreted
the high energy asymptotic behaviour of the  $\gamma_{opt}(\omega)$
(see Fig.\ \ref{fig1}) in terms of its step-like effective counterpart 
 $\gamma_{eff}(\omega)$
(\ref{nel}, \ref{gamefT0}).  As the result, they  arrived at
the spectral function  having a long  tail  up to the  very high
 frequency (see Fig.\ \ref{fig3}d, inset). Nowadays,  similar 
spectral functions occasionally arise in  different models, since this
 shape suggests the spin fluctuations spectrum reported by neutron spectroscopy.   

The calculations was performed in the same fashion as for model 
spectral function shown in Fig.\ \ref{fig3}a.
The coupling constant was chosen 
to reproduce  the value of the {\it tunnelling} gap $\Delta_{tun}$=200 cm$^{-1}$.
The values of the plasma frequency $\omega_{pl}$
and low-frequency value of the non-metallic  part of the dielectric
permeability $\epsilon_{\infty}$ were obtained by the least-square fit.
The peak  position had the  value 330 cm$^{-1}$ which  was close to the one
 in the original paper \cite{collins87}. 
 I choose the power law $1/\omega^{2.5}$ 
for the tail and the high energy cut-off 1600 cm$^{-1}$. 

Let us compare the calculated curve and the experimental one.
Since the low frequency bound in the model spectrum $\Omega_{min}\approx 200$ cm$^{-1}$,
the absorption edge takes  place at  $\Omega_{min}+\Delta_{tun}$ in accord with the 
results discussed in \cite{kamaras88}. At second, in the region between 400 cm$^{-1}$
 and 800 cm$^{-1}$, where the reflectivity ''integrates`` the spectral function,
there is a systematic shift on 140 cm$^{-1}$. For the compensation of this 
shift one has to
decrease the value of tunnelling gap by a. factor of two 
The most important 	discrepancy is connected with the upper frequency bound 
of the single-particle relaxation, that is, with the  $\Omega_{max}$. The 
reflectivity ratio $R_s/R_n$ has a minimum at 1400  cm$^{-1}$ in comparison with 
the 900  cm$^{-1}$ in the experimental one. 
	
In conclusion, the separable {\it d}-wave model reproduce all important 
features
of the traditional ISB  {\it s}-wave one with the single exception:  
 the resulting   images of the spectral function are  
  the convolution of the input 
electron-boson interaction function 
 and the Gauss distribution exp($-\omega^2/\Delta^2$).

\section{Acknowledgements}
 I express my deep gratitude to S.-L.\  Drechsler, O.\ V.\  Dolgov, E.\ G.\ Maksimov, 
A.\ Golubov, D.\  Rainer,  M.\ Kulic,
N.\ Yu.\ Boldyrev, V.\ M.\ Burlakov, and R.\ S.\ Gonnelli
 for numerous discussions, fruitful collaboration, and generous 
support of this work. 
I am deeply grateful to 
 K.D.\ Schotte, K.F.\ Renk, E.A.\ Vinogradov, V.N.\ Agranovich, 
J.\ Fink, and 
V.L.\ Ginzburg for their stimulate decisive support during my work.

The author acknowledges gratefully for financial support of  SFB 311 and 463.

\end{document}